\documentclass[preprint,prd,aps,showpacs,showkeys,nofootinbib]{revtex4}
\usepackage{graphicx}
\begin{document}


\title{Lepton-flavor violation and two loop electroweak corrections to $(g-2)_\mu$ in the B-L symmetry SSM}

\author{Jin-Lei Yang$^{a}$\footnote{JLYangJL@163.com},
Tai-Fu Feng$^{a}$\footnote{fengtf@hbu.edu.cn}, Yu-Li Yan$^{a}$, Wei Li$^{a}$\footnote{WATliwei@163.com}, Shu-Min Zhao$^{a}$, Hai-Bin Zhang$^{a}$\footnote{hbzhang@hbu.edu.cn}}

\affiliation{$^a$Department of Physics, Hebei University, Baoding, 071002, China}

\begin{abstract}
Charged lepton flavor violating processes are forbidden in the standard model (SM), hence the observation of charged lepton flavor transitions would represent a clear signal of new physics beyond the standard model. In this work, we investigate some lepton flavor violating processes in the minimal supersymmetric extension of the SM with local $B-L$ gauge symmetry (B-LSSM). And including the corrections from some two loop diagrams to the anomalous dipole moments (MDM) of muon, we discuss the corresponding constraint on the relevant parameter space of the model. Considering the constraints from updated experimental data, the numerical results show that, new contributions in the B-LSSM enhance the MSSM predictions on the rates of $l_j-l_i$ transitions about one order of magnitude, and also enhance the MSSM prediction on the muon MDM. In addition, two loop electroweak corrections can make important contributions to the muon MDM in the B-LSSM.

\end{abstract}


\maketitle

\section{Introduction\label{sec1}}
\indent\indent

Lepton-flavor-violation (LFV), if observed in the future experiments, is obvious evidence of new physics beyond the standard model (SM), because the lepton-flavor number is conserved in the SM. A detailed analysis of the LFV processes will reveal some properties of high-energy physics, because the processes do not suffer from a large ambiguity due to hadronic matrix elements. In Table~\ref{tab1}, we show the present experimental limits and future sensitivities for the LFV processes $l_j^-\rightarrow l_i^-\gamma,\;l_j^-\rightarrow l_i^-l_i^-l_i^+$~\cite{Adam:2013mnn,Baldini:2013ke,Aubert:2009ag,Hayasaka:2013dsa,Bellgardt:1987du,Blondel:2013ia,Hayasaka:2010np}.
Several predictions for these LFV processes have obtained in the framework of various SM extensions\cite{Ilakovac:1994kj,Diaz:2000cm,Kakizaki:2003jk,Arganda:2005ji,Toma:2013zsa,Zhang:2014osa,Zhao:2015dna}. In this work, we analyze these LFV processes in the minimal supersymmetric extension of the SM with local $B-L$ gauge symmetry ($B-L$ SSM). In addition, it is well known that the magnetic dipole moment (MDM) of muon has close relation with the new physics (NP) beyond the SM, and researching the muon MDM is an effective way to find NP beyond the SM. Hence including some two loop diagrams, we also analyze the muon MDM in the $B-L$ SSM.

The $B-L$ SSM~\cite{5,6,addP1,addP2} is based on the gauge symmetry group $SU(3)_C\otimes SU(2)_L\otimes U(1)_Y\otimes U(1)_{B-L}$, where $B$ stands for the baryon number and $L$ stands for the lepton number respectively. Besides accounting elegantly for the existence and smallness of the left-handed neutrino masses, the $B-L$ SSM also alleviates the little hierarchy problem of the MSSM~\cite{search}, because the exotic singlet Higgs and right-handed (s)neutrinos~\cite{7,77,88,9,99,10,11} release additional parameter space from the LHC constraints. The invariance under $U(1)_{B-L}$ gauge group imposes the R-parity conservation which is assumed in the MSSM to avoid proton decay. And R-parity conservation can be maintained if $U(1)_{B-L}$ symmetry is broken spontaneously~\cite{C.S.A}. Furthermore, the model can provide much more candidates for the dark matter comparing that with the MSSM~\cite{16,1616,DelleRose:2017ukx,DelleRose:2017uas}.

The paper is organized as follows. In Sec.II, the main ingredients of the $B-L$ SSM are summarized briefly by introducing the superpotential and the general soft breaking terms. We present the analysis on the decay width of the rare LFV processes and the muon MDM in Sec.III. The numerical analyses are given in Sec.IV, and Sec.V gives a summary. The tedious formulae are collected in Appendices.

\begin{table*}
\begin{tabular*}{\textwidth}{@{\extracolsep{\fill}}lll@{}}
\hline
LFV process & Present limit & Future sensitivity\\
\hline
$\mu\rightarrow e\gamma$ & $<5.7\times10^{-13}$\cite{Adam:2013mnn} & $\sim6\times10^{-14}$\cite{Baldini:2013ke}\\
$\mu\rightarrow 3e$ & $<1\times10^{-12}$\cite{Bellgardt:1987du} & $\sim10^{-16}$\cite{Blondel:2013ia}\\
$\tau\rightarrow e\gamma$ & $<3.3\times10^{-8}$\cite{Aubert:2009ag} & $\sim10^{-8}-10^{-9}$\cite{Hayasaka:2013dsa}\\
$\tau\rightarrow 3e$ & $<2.7\times10^{-8}$\cite{Hayasaka:2010np} & $\sim10^{-9}-10^{-10}$\cite{Hayasaka:2013dsa}\\
$\tau\rightarrow \mu\gamma$ & $<4.4\times10^{-8}$\cite{Aubert:2009ag} & $\sim10^{-8}-10^{-9}$\cite{Hayasaka:2013dsa}\\
$\tau\rightarrow 3\mu$ & $<2.1\times10^{-8}$\cite{Hayasaka:2010np} & $\sim10^{-9}-10^{-10}$\cite{Hayasaka:2013dsa}\\
\hline
\end{tabular*}
\caption{Present limits and future sensitivities for the branching ratios for the LFV processes.}
\label{tab1}
\end{table*}

\section{The $B-L$ SSM\label{sec2}}
\indent\indent
In literatures there are several popular versions of $B-L$ SSM. Here we adopt the version described in Refs.~\cite{44,Abdallah:2014fra,8,Khalil:2015wua} to proceed our analysis, which allows for a spontaneously broken $U(1)_{B-L}$ without necessarily breaking R-parity. This requires the addition of two chiral singlet superfields $\hat{\eta}_{1}\sim(1,1,0,-1)$, $\hat{\eta}_{2}\sim(1,1,0,1)$, as well as three generations of right-handed neutrinos. In addition, this version of $B-L$ SSM is encoded in SARAH~\cite{164,165,166,167,168} which is used to create the mass matrices and interaction vertexes in the model. Meanwhile, quantum numbers of the matter chiral superfields for quarks and leptons are given by
\begin{eqnarray}
&&\hat{Q}_i=\left(\begin{array}{c}\hat u_i\\ \hat d_i\end{array}\right)\sim(3, 2, 1/6, 1/6), \quad\;\hat{L}_i=\left(\begin{array}{c}\hat \nu_i\\ \hat e_i\end{array}\right)\sim(1, 2, -1/2, -1/2),\nonumber\\
&&\hat{U}^c_i\sim(3, 1, -2/3, -1/6),\quad\; \hat{D}^c_i\sim(3, 1, 1/3, -1/6), \quad\;\hat{E}^c_i\sim(1, 1, 1, 1/2),
\end{eqnarray}
with $i=1,2,3$ denoting the index of generation. In addition, the quantum numbers of two Higgs doublets are assigned as
\begin{eqnarray}
&&\hat{H_1}=\left(\begin{array}{c}H_1^1\\ H_1^2\end{array}\right)\sim (1, 2, -1/2, 0),\quad\;\hat{H_2}=\left(\begin{array}{c}H_2^1\\ H_2^2\end{array}\right)\sim(1, 2, 1/2, 0).
\end{eqnarray}

The corresponding superpotential of the $B-L$ SSM is written as
\begin{eqnarray}
&&W=Y_u^{ij}\hat{Q_i}\hat{H_2}\hat{U_j^c}+\mu \hat{H_1} \hat{H_2}-Y_d^{ij} \hat{Q_i} \hat{H_1} \hat{D_j^c}
-Y_e^{ij} \hat{L_i} \hat{H_1} \hat{E_j^c}+\nonumber\\
&&\;\;\;\;\;\;\;\;\;Y_{\nu, ij}\hat{L_i}\hat{H_2}\hat{\nu}^c_j-\mu' \hat{\eta}_1 \hat{\eta}_2
+Y_{x, ij} \hat{\nu}_i^c \hat{\eta}_1 \hat{\nu}_j^c,
\end{eqnarray}
where $i, j$ are generation indices. Correspondingly, the soft breaking terms of the $B-L$ SSM are generally given as
\begin{eqnarray}
&&\mathcal{L}_{soft}=\Big[-\frac{1}{2}(M_1\tilde{\lambda}_{B} \tilde{\lambda}_{B}+M_2\tilde{\lambda}_{W} \tilde{\lambda}_{W}+M_3\tilde{\lambda}_{g} \tilde{\lambda}_{g}+2M_{BB^{'}}\tilde{\lambda}_{B^{'}} \tilde{\lambda}_{B}+M_{B^{'}}\tilde{\lambda}_{B^{'}} \tilde{\lambda}_{B^{'}})-
\nonumber\\
&&\hspace{1.4cm}
B_\mu H_1H_2 -B_{\mu^{'}}\tilde{\eta}_1 \tilde{\eta}_2 +T_{u,ij}\tilde{Q}_i\tilde{u}_j^cH_2+T_{d,ij}\tilde{Q}_i\tilde{d}_j^cH_1+
T_{e,ij}\tilde{L}_i\tilde{e}_j^cH_1+T_{\nu}^{ij} H_2 \tilde{\nu}_i^c \tilde{L}_j+\nonumber\\
&&\hspace{1.4cm}
T_x^{ij} \tilde{\eta}_1 \tilde{\nu}_i^c \tilde{\nu}_j^c+h.c.\Big]-m_{\tilde{\nu},ij}^2(\tilde{\nu}_i^c)^* \tilde{\nu}_j^c-
m_{\tilde{q},ij}^2\tilde{Q}_i^*\tilde{Q}_j-m_{\tilde{u},ij}^2(\tilde{u}_i^c)^*\tilde{u}_j^c-m_{\tilde{\eta}_1}^2 |\tilde{\eta}_1|^2-\nonumber\\
&&\hspace{1.4cm}
m_{\tilde{\eta}_2}^2 |\tilde{\eta}_2|^2-m_{\tilde{d},ij}^2(\tilde{d}_i^c)^*\tilde{d}_j^c-m_{\tilde{L},ij}^2\tilde{L}_i^*\tilde{L}_j-
m_{\tilde{e},ij}^2(\tilde{e}_i^c)^*\tilde{e}_j^c-m_{H_1}^2|H_1|^2-m_{H_2}^2|H_2|^2,
\end{eqnarray}
with $\lambda_{B}, \lambda_{B^{'}}$ denoting the gaugino of $U(1)_Y$ and $U(1)_{(B-L)}$ respectively. The local gauge symmetry $SU(2)_L\otimes U(1)_Y\otimes U(1)_{B-L}$ breaks down to the electromagnetic symmetry $U(1)_{em}$ as the Higgs fields receive vacuum expectation values:
\begin{eqnarray}
&&H_1^1=\frac{1}{\sqrt2}(v_1+{\rm Re}H_1^1+i{\rm Im}H_1^1),
\qquad\; H_2^2=\frac{1}{\sqrt2}(v_2+{\rm Re}H_2^2+i{\rm Im}H_2^2),\nonumber\\
&&\tilde{\eta}_1=\frac{1}{\sqrt2}(u_1+{\rm Re}\tilde{\eta}_1+i{\rm Im}\tilde{\eta}_1),
\qquad\;\quad\;\tilde{\eta}_2=\frac{1}{\sqrt2}(u_2+i{\rm Re}\tilde{\eta}_2+i{\rm Im}\tilde{\eta}_2)\;.
\end{eqnarray}
For convenience, we define $u^2=u_1^2+u_2^2,\; v^2=v_1^2+v_2^2$ and $\tan\beta^{'}=\frac{u_2}{u_1}$ in analogy to the ratio of the MSSM VEVs ($\tan\beta=\frac{v_2}{v_1}$).

The presence of two Abelian groups gives rise to a new effect absent in the MSSM or other SUSY models with just one Abelian gauge group: the gauge kinetic mixing. This mixing couples the $B-L$ sector to the MSSM sector, and even if it is set to zero at $M_{GUT}$, it can be induced through RGEs\cite{RGE1,RGE2,RGE3,RGE4,RGE5,RGE6,RGE7}.  In practice, it turns out that it is easier to work with non-canonical covariant derivatives instead of off-diagonal field-strength tensors. However, both approaches are equivalent\cite{R.F}. Hence in the following, we consider covariant derivatives of the form
\begin{eqnarray}
&&D_\mu=\partial_\mu-i\left(\begin{array}{cc}Y,&B-L\end{array}\right)
\left(\begin{array}{cc}g_{_Y},&g_{_{YB}}^{'}\\g_{_{BY}}^{'},&g_{_{B-L}}\end{array}\right)
\left(\begin{array}{c}A_{_\mu}^{\prime Y} \\ A_{_\mu}^{\prime BL}\end{array}\right)\;,
\label{gauge1}
\end{eqnarray}
where $A_{_\mu}^{\prime Y}, A^{\prime\mu, BL}$ denote the gauge fields associated with the two $U(1)$ gauge groups, $Y, B-L$ corresponding to the hypercharge and $B-L$ charge respectively. As long as the two Abelian gauge groups are unbroken, we still have the freedom to perform a change of the basis. Choosing $R$ in a proper form, one can write the coupling matrix as
\begin{eqnarray}
&&\left(\begin{array}{cc}g_{_Y},&g_{_{YB}}^{'}\\g_{_{BY}}^{'},&g_{_{B-L}}\end{array}\right)
R^T=\left(\begin{array}{cc}g_{_1},&g_{_{YB}}\\0,&g_{_{B}}\end{array}\right)\;,
\label{gauge3}
\end{eqnarray}
where $g_{_{1}}$ corresponds to the measured hypercharge coupling which is modified in
$B-L$ SSM as given along with $g_{_{B}}$ and $g_{_{YB}}$ in Refs.~\cite{BLSSM1}. Then, we can redefine the $U(1)$ gauge fields
\begin{eqnarray}
&&R\left(\begin{array}{c}A_{_\mu}^{\prime Y} \\ A_{_\mu}^{\prime BL}\end{array}\right)
=\left(\begin{array}{c}A_{_\mu}^{Y} \\ A_{_\mu}^{BL}\end{array}\right)\;.
\label{gauge4}
\end{eqnarray}

Immediate interesting consequence of the gauge kinetic mixing arise in  various sectors of the model as discussed in the subsequent analysis. First, $A^{BL}$ boson mixes at the tree level with the $A^Y$ and $V^3$ bosons. In the basis $(A^Y, V^3, A^{BL})$, the corresponding mass matrix reads,
\begin{eqnarray}
&&\left(\begin{array}{*{20}{c}}
\frac{1}{8}g_{_1}^2 v^2 & -\frac{1}{8}g_{_1}g_{_2} v^2 & \frac{1}{8}g_{_1}g_{_{YB}} v^2 \\ [6pt]
-\frac{1}{8}g_{_1}g_{_2} v^2 & \frac{1}{8}g_{_2}^2 v^2 & -\frac{1}{8}g_{_2}g_{_{YB}} v^2\\ [6pt]
\frac{1}{8}g_{_1}g_{_{YB}} v^2 & -\frac{1}{8}g_{_2}g_{_{YB}} v^2 & \frac{1}{8}g_{_{YB}}^2 v^2+\frac{1}{8}g_{_{B}}^2 u^2
\end{array}\right).\label{gauge matrix}
\end{eqnarray}
This mass matrix can be diagonalized by a unitary mixing matrix, which can be expressed by two mixing angles $\theta_{_W}$ and $\theta_{_W}'$ as
\begin{eqnarray}
&&\left(\begin{array}{*{20}{c}}
\gamma\\ [6pt]
Z\\ [6pt]
Z'
\end{array}\right)=
\left(\begin{array}{*{20}{c}}
\cos\theta_{_W} & \sin\theta_{_W} & 0 \\ [6pt]
-\sin\theta_{_W}\cos\theta_{_W}' & \cos\theta_{_W}\cos\theta_{_W}' & \sin\theta_{_W}'\\ [6pt]
\sin\theta_{_W}\sin\theta_{_W}' & -\cos\theta_{_W}'\sin\theta_{_W}' & \cos\theta_{_W}'
\end{array}\right)
\left(\begin{array}{*{20}{c}}
A^Y\\ [6pt]
V^3\\ [6pt]
A^{BL}
\end{array}\right).
\end{eqnarray}
Then $\sin^2\theta_{_W}'$ can be written as
\begin{eqnarray}
\sin^2\theta_{_W}'=\frac{1}{2}-\frac{(g_{_{YB}}^2-g_{_1}^2-g_{_2}^2)x^2+
4g_{_B}^2}{2\sqrt{(g_{_{YB}}^2+g_{_1}^2+g_{_2}^2)x^4+8g_{_B}^2(g_{_{YB}}^2-g_{_1}^2-g_{_2}^2)x^2+16g_{_B}^2}},
\end{eqnarray}
where $x=\frac{v}{u}$. Compared with the MSSM, this $Z-Z'$ mixing makes new contributions to the $l_j\rightarrow 3l_i$ decay channel, and the new mixing angle $\theta_{W'}$ appears in the couplings involve $Z$ boson. The exact eigenvalues of Eq.(\ref{gauge matrix}) are given by
\begin{eqnarray}
&&\qquad\;\quad\;m_\gamma^2=0,\nonumber\\
&&\qquad\;\quad\;m_{Z,{Z^{'}}}^2=\frac{1}{8}\Big((g_{_1}^2+g_2^2+g_{_{YB}}^2)v^2+4g_{_B}^2u^2 \nonumber\\
&&\qquad\;\qquad\;\qquad\;\mp\sqrt{(g_{_1}^2+g_{_2}^2+g_{_{YB}}^2)^2v^4+8(g_{_{YB}}^2-g_{_1}^2-
g_{_2}^2)g_{_B}^2v^2u^2+16g_{_B}^4u^4}\Big),
\end{eqnarray}

Then the gauge kinetic mixing leads to the mixing between the $H_1^1,\;H_2^2,\;\tilde{\eta}_1,\;\tilde{\eta}_2$ at the tree level. In the basis (${\rm Re}H_1^1$, ${\rm Re}H_2^2$, ${\rm Re}\tilde{\eta}_1$, ${\rm Re}\tilde{\eta}_2$),
the tree level mass squared matrix for scalar Higgs bosons is given by
\begin{eqnarray}
&&M_h^2=u^2\times\nonumber\\
&&\left(\begin{array}{*{20}{c}}
{\frac{1}{4}\frac{g^2 x^2}{1+\tan\beta^2}+n^2\tan\beta}&{-\frac{1}{4}g^2\frac{x^2\tan\beta}{1+\tan^2\beta}}-n^2&
{\frac{1}{2}g_{_B}g_{_{YB}}\frac{x}{T}}&
{-\frac{1}{2}g_{_B}g_{_{YB}}\frac{x\tan\beta'}{T}}\\ [6pt]
{-\frac{1}{4}g^2\frac{ x^2\tan\beta}{1+\tan^2\beta}}-n^2&{\frac{1}{4}\frac{g^2\tan^2\beta x^2}{1+\tan\beta^2}+\frac{n^2}{\tan\beta}}&
{\frac{1}{2}g_{_B}g_{_{YB}}\frac{x\tan\beta}{T}}&{\frac{1}{2}g_{_B}g_{_{YB}}\frac{x\tan\beta\tan\beta'}{T}}\\ [6pt]
{\frac{1}{2}g_{_B}g_{_{YB}}\frac{x}{T}}&{\frac{1}{2}g_{_B}g_{_{YB}}\frac{x\tan\beta}{T}}&{\frac{g_{_B}^2}{1+\tan^2\beta'}+\tan\beta'N^2}&
{-g_{_B}^2\frac{\tan\beta'}{1+\tan^2\beta'}-N^2}\\ [6pt]
{-\frac{1}{2}g_{_B}g_{_{YB}}\frac{x\tan\beta'}{T}}&{\frac{1}{2}g_{_B}g_{_{YB}}\frac{x\tan\beta\tan\beta'}{T}}&
{-g_{_B}^2\frac{\tan\beta'}{1+\tan^2\beta^{'}}-N^2}&{g_{_B}^2\frac{\tan^2\beta'}{1+\tan^2\beta'}+\frac{N^2}{\tan\beta'}}
\end{array}\right)
\end{eqnarray}
where $g^2=g_{_1}^2+g_{_2}^2+g_{_{YB}}^2$, $T=\sqrt{1+\tan^2\beta}\sqrt{1+\tan^2\beta'}$,
$n^2=\frac{{\rm Re}B\mu}{u^2}$ and $N^2=\frac{{\rm Re}B\mu^{'}}{u^2}$, respectively. Compared the MSSM, this new mixing in the $B-L$ SSM can affect the following analysis.

Including the leading-log radiative corrections from stop and top particles~\cite{HiggsC1,HiggsC2,HiggsC3}, the mass of the SM-like Higgs boson can be written as
\begin{eqnarray}
&&m_h=\sqrt{(m_{h_1}^0)^2+\Delta m_h^2},\label{higgs mass}
\end{eqnarray}
where $m_{h_1}^0$ denotes the lightest tree-level Higgs boson mass, and
\begin{eqnarray}
&&\Delta m_h^2=\frac{3m_t^4}{2\pi v^2}\Big[\Big(\tilde{t}+\frac{1}{2}+\tilde{X}_t\Big)+\frac{1}{16\pi^2}\Big(\frac{3m_t^2}{2v^2}-32\pi\alpha_3\Big)\Big(\tilde{t}^2
+\tilde{X}_t \tilde{t}\Big)\Big],\nonumber\\
&&\tilde{t}=log\frac{M_S^2}{m_t^2},\qquad\;\tilde{X}_t=\frac{2\tilde{A}_t^2}{M_S^2}\Big(1-\frac{\tilde{A}_t^2}{12M_S^2}\Big),\label{higgs corrections}
\end{eqnarray}
here $\alpha_3$ is the strong coupling constant, $M_S=\sqrt{m_{\tilde t_1}m_{\tilde t_2}}$ with $m_{\tilde t_{1,2}}$ denoting the stop masses, $\tilde{A}_t=A_t-\mu \cot\beta$ with $A_t=T_{u,33}$ being the trilinear Higgs stop coupling and $\mu$ denoting the Higgsino mass parameter.

Meanwhile, additional D-terms contribute to the mass matrices of the squarks and sleptons, and sleptons also affect the subsequent analysis. On the basis $(\tilde L, \tilde e^c)$, the mass matrix of sleptons can be written as
\begin{eqnarray}
&&m_{\tilde e}^2=
\left(\begin{array}{cc}m_{eL}^2,&\frac{1}{\sqrt2}(v_1 T_e^\dagger-v_2\mu Y_e^\dagger)\\\frac{1}{\sqrt2}(v_1 T_e-v_2\mu^* Y_e),&m_{eR}^2\end{array}\right),
\end{eqnarray}
\begin{eqnarray}
&&m_{eL}^2=\frac{1}{8}\Big[2g_{_B}(g_{_B}+g_{_{YB}})(u_1^2-u_2^2)+(g_1^2-g_2^2+g_{_{YB}}^2+2g_{_B}g_{_{YB}})(v_1^2-
v_2^2)\Big]\nonumber\\
&&\qquad\;\quad\;+m_{\tilde L}^2+\frac{v_1^2}{2}Y_e^\dagger Y_e,\nonumber\\
&&m_{eR}^2=\frac{1}{24}\Big[2g_{_B}(g_{_B}+2g_{_{YB}})(u_2^2-u_1^2)+2(g_1^2+g_{_{YB}}^2+2g_{_B}g_{_{YB}})(v_2^2-
v_1^2)\Big]\nonumber\\
&&\qquad\;\quad\;+m_{\tilde e}^2+\frac{v_1^2}{2}Y_e^\dagger Y_e.\label{eq17}
\end{eqnarray}
It can be noted that $\tan\beta'$ and new gauge coupling constants $g_{_B}$, $g_{_{YB}}$ in the $B-L$ SSM can affect the mass matrix of sleptons.


\section{Lepton flavor violation and $(g-2)_\mu$ in the $B-L$ SSM\label{sec3}}
\indent\indent
In this section, we present the analysis on the decay width of the rare LFV processes $l_j^-\rightarrow l_i^-\gamma$ and $l_j^-\rightarrow l_i^-l_i^-l_i^+$ in the $B-L$ SSM. In addition, considering the corrections from some two loop diagrams, we analyze the NP contributions to the muon MDM, $\Delta a_\mu^{NP}$.

\subsection{Rare decay $l_j^-\rightarrow l_i^-\gamma$}
\begin{figure}
\setlength{\unitlength}{1mm}
\centering
\includegraphics[width=5in]{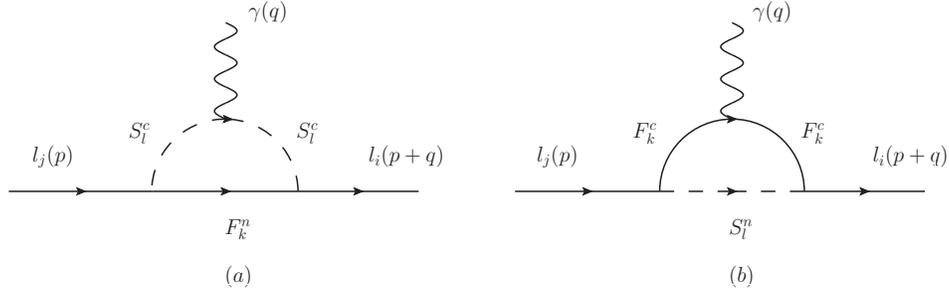}
\vspace{0cm}
\caption[]{Feynman diagrams for the process $l_j^-\rightarrow l_i^-\gamma$. (a) represents the contributions from neutral fermions $F^n_k$ and charge scalars loops $S^c_l$, and (b) represents the contributions from charged fermions $F^c_k$ and neutral scalars or pseudoscalars $S^n_l$ loops.}
\label{figllr}
\end{figure}

At first, the off-shell amplitude for $l_j^-\rightarrow l_i^-\gamma$ is generally written as\cite{Hisano:1995cp}
\begin{eqnarray}
&&T=e\epsilon^\mu \bar u_i(p+q)[q^2 \gamma_\mu(A_1^LP_L+A_1^RP_R)+m_{l_j}i\sigma_{\mu\nu}q^\nu(A_2^LP_L+A_2^RP_R)]u_j(p),\label{Allr}
\end{eqnarray}
in the limit $q\rightarrow0$ with $q$ being photon momentum. In addition, $p$ is the momentum of the particle $l_j$, $\epsilon$ is the photon polarization vector, $u_i$ (and $\nu_i$ in the expression below) is the wave function for lepton (antilepton), and $P_L=(1-\gamma_5)/2$, $P_R=(1+\gamma_5)/2$. Then, the Feynman diagrams contributing to the above amplitude are depicted by Fig.~\ref{figllr}.
Calculating the Feynman diagrams, the coefficients $A_{1,2}^{L,R}$ in Eq.(\ref{Allr}) can be written as
\begin{eqnarray}
&&A_1^{L,R}=A_1^{(a)L,R}+A_1^{(b)L,R},\nonumber\\
&&A_2^{L,R}=A_2^{(a)L,R}+A_2^{(b)L,R},
\end{eqnarray}
where the concrete expressions for $A_{1,2}^{(a)L,R}$, $A_{1,2}^{(b)L,R}$ corresponding to Fig.~\ref{figllr}(a), (b) can be found in Appendix \ref{wilsonllr}.

Using the amplitude Eq.(\ref{Allr}), the decay width for $l_j^-\rightarrow l_i^-\gamma$ can be obtained easily as
\begin{eqnarray}
&&\Gamma(l_j^-\rightarrow l_i^-\gamma)=\frac{e^2}{16\pi}m_{l_j}^5(|A_2^L|^2+|A_2^R|^2).
\end{eqnarray}
And the branching ratio for $l_j^-\rightarrow l_i^-\gamma$ is
\begin{eqnarray}
&&Br(l_j^-\rightarrow l_i^-\gamma)=\frac{\Gamma(l_j^-\rightarrow l_i^-\gamma)}{\Gamma_{l_j^-}},
\end{eqnarray}
where $\Gamma_{l_j^-}$ is the total decay width of the lepton $l_j^-$. In the numerical calculation, we use $\Gamma_\mu\approx2.996\times10^{-19}{\rm GeV}$ for the muon and $\Gamma_\tau\approx2.265\times10^{-12}{\rm GeV}$ for the tauon\cite{PDG}.

\subsection{Rare decay $l_j^-\rightarrow l_i^-l_i^-l_i^+$}
\begin{figure}
\setlength{\unitlength}{1mm}
\centering
\includegraphics[width=2in]{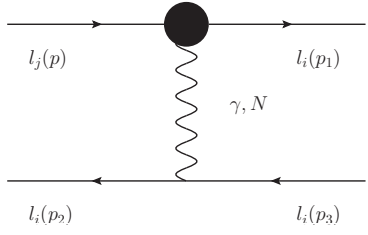}
\vspace{0cm}
\caption[]{Penguin-type diagram for the process $l_j^-\rightarrow l_i^-l_i^-l_i^+$. The blob indicates an $l_j^-l_i^-\gamma$ vertex such as Fig.~\ref{figllr} or $l_j^-l_i^-N$ vertex where $N$ denotes $Z$ and $Z'$ bosons.}
\label{figl3lPenguin}
\end{figure}
For the process $l_j^-\rightarrow l_i^-l_i^-l_i^+$, the effective amplitude includes the contributions from penguin-type and box-type diagrams. Using Eq.(\ref{Allr}), the contributions from $\gamma$-penguin diagrams can be written as\cite{Hisano:1995cp}
\begin{eqnarray}
&&T_{\gamma-{\rm penguin}}=\bar u_i(p_1)[q^2 \gamma_\mu(A_1^LP_L+A_1^RP_R)+m_{l_j}i\sigma_{\mu\nu}q^\nu(A_2^LP_L+A_2^RP_R)]u_j(p)\nonumber\\
&&\qquad\qquad\quad\times\frac{e^2}{q^2}\bar u_i(p_2)\gamma^\mu \nu_i(p_3)-(p_1\leftrightarrow p_2),\label{rpenguin}
\end{eqnarray}
As shown in Fig.\ref{figl3lPenguin}, the contribution from $N$-penguin (here $N$ represents $Z$ and $Z'$ bosons) diagrams can be written as\cite{Hisano:1995cp}
\begin{eqnarray}
&&T_{N-{\rm penguin}}=\frac{e^2}{m_N^2}\bar u_i(p_1)\gamma_\mu(F^LP_L+F^RP_R)u_j(p)\bar u_i(p_2)\gamma^\mu(C_{\bar l_iNl_i}^LP_L\nonumber\\
&&\qquad\qquad\quad+C_{\bar l_iNl_i}^RP_R)\nu_i(p_3)-(p_1\leftrightarrow p_2),\label{Npenguin}
\end{eqnarray}
where $m_N$ denotes the mass for $Z$ or $Z'$ boson, and the concrete expressions for $F^{L,R}$ are given in Appendix \ref{wilsonl3l}. There is also dipole term for $Z$ and $Z'$ contributions, but we neglect it in the calculation. Because $\gamma$ exchanges in Fig. 2 represent the electromagnetic interaction, the corresponding gauge symmetry is not broken, while $Z$ or $Z'$ exchanges denote the weak interaction, the corresponding gauge symmetry is broken. And $q$ is the sum of the momentum of two outward on-shell lepton, which is about the same order of magnitude as that of $m_e$ or $m_\mu$. So we cannot neglect it for $\gamma$ contributions in Eq.(\ref{rpenguin}), because photon is massless. But $q$ is negligible compared with $m_N$, hence we neglect it for $Z$ or $Z'$ contributions in Eq.(\ref{Npenguin}).

\begin{figure}
\setlength{\unitlength}{1mm}
\centering
\includegraphics[width=4in]{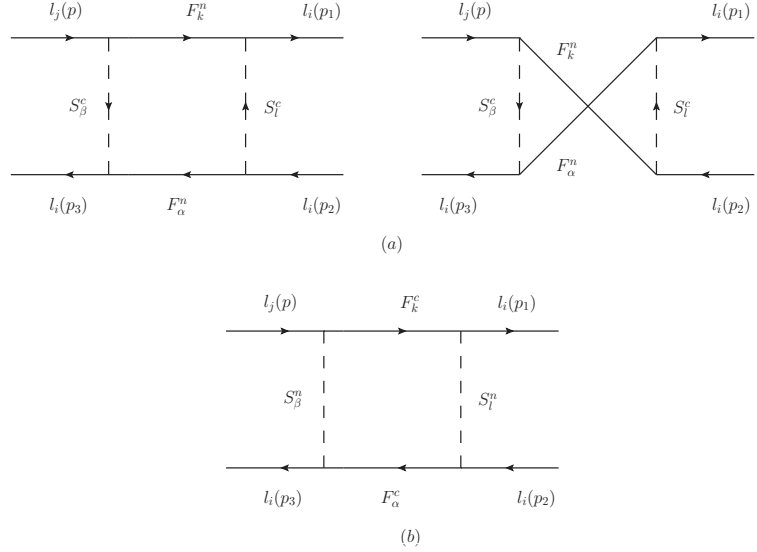}
\vspace{0cm}
\caption[]{Box-type diagrams for the process $l_j^-\rightarrow l_i^-l_i^-l_i^+$. (a) represents the contributions from neutral fermions $F^n_{k,\alpha}$ and charge scalars loops $S^c_{l,\beta}$, and (b) represents the contributions from charged fermions $F^c_{k,\alpha}$ and neutral scalars or pseudoscalars $S^n_{l,\beta}$ loops.}
\label{figl3lbox}
\end{figure}

In addition, box-type diagrams can also contribute to the process $l_j^-\rightarrow l_i^-l_i^-l_i^+$. The corresponding Feynman diagrams are drawn in Fig.\ref{figl3lbox}.
Then the effective Hamiltonian for the box-type diagrams can be written as
\begin{eqnarray}
&&T_{box}=\Big\{B_1^Le^2\bar u_i(p_1)\gamma_\mu P_Lu_j(p)\bar u_i(p_2)\gamma^\mu P_L\nu_i(p_3)+(L\leftrightarrow R)\Big\}\nonumber\\
&&\qquad\quad +\Big\{B_2^L[e^2\bar u_i(p_1)\gamma_\mu P_Lu_j(p)\bar u_i(p_2)\gamma^\mu P_R\nu_i(p_3)-(p_1\leftrightarrow p_2)]+(L\leftrightarrow R)\Big\}\nonumber\\
&&\qquad\quad +\Big\{B_3^L[e^2\bar u_i(p_1)P_Lu_j(p)\bar u_i(p_2)P_R\nu_i(p_3)-(p_1\leftrightarrow p_2)]+(L\leftrightarrow R)\Big\}\nonumber\\
&&\qquad\quad +\Big\{B_4^L[e^2\bar u_i(p_1)\sigma_{\mu,\nu}P_Lu_j(p)\bar u_i(p_2)\sigma^{\mu,\nu}P_L\nu_i(p_3)-(p_1\leftrightarrow p_2)]+(L\leftrightarrow R)\Big\},
\end{eqnarray}
where the coefficients $B_{1,2,3,4}^{L,R}$ originate from those box diagrams in Fig.\ref{figl3lbox}, and the concrete expressions can be found in Appendix \ref{wilsonl3l}. Then the decay width for $l_j^-\rightarrow l_i^-l_i^-l_i^+$ are\cite{Hisano:1995cp}
\begin{eqnarray}
&&\Gamma(l_j^-\rightarrow l_i^-l_i^-l_i^+)=\frac{e^4m_{l_j}^5}{512\pi^3}\Big\{(|A_2^L|^2+|A_2^R|^2)\Big(\frac{16}{3}\ln\frac{m_{l_j}}{2m_{l_i}}-
\frac{14}{9}\Big)\nonumber\\
&&\qquad\quad\qquad\quad\qquad+(|A_1^L|^2+|A_1^R|^2)-2(A_1^LA_2^{R*}+A_2^LA_1^{R*}+H.c.)+\frac{1}{6}(|B_1^L|^2+|B_1^R|^2)
\nonumber\\
&&\qquad\quad\qquad\quad\qquad+\frac{1}{3}(|B_2^L|^2+|B_2^R|^2)+\frac{1}{24}(|B_3^L|^2+|B_3^R|^2)+6(|B_4^L|^2+|B_4^R|^2)
\nonumber\\
&&\qquad\quad\qquad\quad\qquad-\frac{1}{2}(B_3^LB_4^{L*}+B_3^RB_4^{R*}+H.c.)+\frac{1}{3}(A_1^LB_1^{L*}+A_1^RB_1^{R*}+
A_1^LB_2^{L*}\nonumber\\
&&\qquad\quad\qquad\quad\qquad+A_1^RB_2^{R*}+H.c.)-\frac{2}{3}(A_2^RB_1^{L*}+A_2^LB_1^{R*}+A_2^LB_2^{R*}+A_2^RB_2^{R*}+
H.c.)\nonumber\\
&&\qquad\quad\qquad\quad\qquad+\frac{1}{3}\Big[2(|F^{LL}|^2+|F^{RR}|^2)+(|F^{LR}|^2+|F^{RL}|^2)+(B_1^LF^{LL*}+
B_1^RF^{RR*}\nonumber\\
&&\qquad\quad\qquad\quad\qquad+B_2^LF^{LR*}+B_2^RF^{RL*}+H.c.)+2(A_1^LF^{LL*}+A_1^RLF^{RR*}+H.c.)\nonumber\\
&&\qquad\quad\qquad\quad\qquad+(A_1^LF^{LR*}+A_1^RLF^{RL*}+H.c.)-4(A_2^RF^{LL*}+A_2^LF^{RR*}+H.c.)\nonumber\\
&&\qquad\quad\qquad\quad\qquad-2(A_2^LF^{RL*}+A_2^RLF^{LR*}+H.c.)\Big]\Big\},
\end{eqnarray}
where
\begin{eqnarray}
&&F^{LL}=\sum_{N=Z,Z'}\frac{F^LC_{\bar l_i Nl_i}^L}{m_N^2},\;\;\;\;\;\;F^{RR}=F^{LL}(L\leftrightarrow R),\nonumber\\
&&F^{LR}=\sum_{N=Z,Z'}\frac{F^LC_{\bar l_i Nl_i}^R}{m_N^2},\;\;\;\;\;\;F^{LR}=F^{RL}(L\leftrightarrow R).
\end{eqnarray}

\subsection{$(g-2)_\mu$}
Finally, we analyze the muon MDM. The difference between experiment and the SM prediction on $a_\mu$ is\cite{Bennett:2006fi,Mohr:2008fa}
\begin{eqnarray}
&&\Delta a_\mu=a_\mu^{exp}-a_\mu^{SM}=(24.8\pm7.9)\times10^{-10},
\label{aulimit}
\end{eqnarray}
where all errors combining in the quadrature. Several predictions for the muon MDM have been discussed in the framework of various SM extensions\cite{Abel:1991dv,Moroi:1995yh,Feng:2001tr,Martin:2001st,Diaz:2002tp,Cheung:2009fc,Zhao:2014dxa,
Feng:2008cn,Feng:2008nm,Feng:2009gn,Yang:2009zzh}. The muon MDM can actually be expressed as the operators
\begin{eqnarray}
&&\mathcal{L}_{MDM}=\frac{e}{4m_\mu}a_\mu\bar l_\mu\sigma^{\alpha\beta}l_\mu F_{\alpha\beta}.
\label{LMDM}
\end{eqnarray}
Here, $\sigma^{\alpha\beta}=i[\gamma^\alpha,\gamma^\beta]/2$, $l_\mu$ represents the wave function for muon, $m_\mu$ is the muon mass, $a_\mu=\frac{1}{2}(g-2)_\mu$ and $F^{\alpha\beta}$ is the electromagnetic field strength. Adopting the effective Lagrangian approach, we can get\cite{Feng:2008cn,Feng:2008nm,Feng:2009gn}
\begin{eqnarray}
&&a_\mu=\frac{4Q_fm_\mu^2}{(4\pi)^2}\Re(C_2^R+C_2^{L*}+C_6^R),
\label{MDM}
\end{eqnarray}
where $Q_f=-1$, and $C_{2,6}^{L,R}$ represent the Wilson coefficients of the corresponding operators $O_{2,6}^{L,R}$
\begin{eqnarray}
&&O_2^{L,R}=\frac{eQ_f}{(4\pi)^2}(-iD_\alpha^*) \bar l_\mu \gamma^\alpha F\cdot\sigma P_{L,R}l_\mu,\nonumber\\
&&O_6^{L,R}=\frac{eQ_fm_\mu}{(4\pi)^2}\bar l_\mu F\cdot\sigma P_{L,R}l_\mu,
\end{eqnarray}
where $D_\alpha=\partial_\alpha+ie A_\alpha$. Then, through the calculation of Fig.~\ref{figllr}, the one loop contributions to the muon MDM can be written as
\begin{eqnarray}
&&a_\mu^{one-loop}=a_\mu^{(a)}+a_\mu^{(b)},
\label{oneloop MDM}
\end{eqnarray}
where $a_\mu^{(a,b)}$ are the contributions to muon MDM corresponding to Fig.~\ref{figllr}(a), (b) respectively, and the concrete expressions of them are given in Appendix \ref{au}.

\begin{figure}
\setlength{\unitlength}{1mm}
\centering
\includegraphics[width=6in]{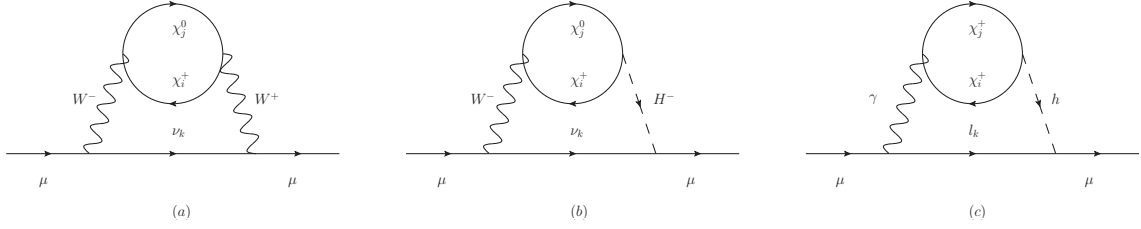}
\vspace{0cm}
\caption[]{The two loop Barr-Zee type diagrams in which a closed fermion loop is attached to the virtual gauge bosons or Higgs fields, the corresponding contributions to the muon MDM are obtained by attaching a photon to the internal particles in all possible ways.}
\label{Barzeediagrams}
\end{figure}

In addition, the two loop Barr-Zee type diagrams can give important contributions to the muon MDM. According to the decoupling theorem, the contributions from the two loop diagrams with a closed slepton loop are suppressed by heavy slepton mass, we neglect these diagrams in the following calculation. Then we consider the contributions from the two loop diagrams in which a closed fermion loop is attached to the virtual gauge bosons or Higgs fields. According to Ref.\cite{Yang:2009zzh}, the main two loop diagrams contributing to the muon MDM are shown in Fig.\ref{Barzeediagrams}.
Then, including the two loop corrections, the muon MDM are given by
\begin{eqnarray}
&&a_\mu^{two-loop}=a_\mu^{one-loop}+a_\mu^{WW}+a_\mu^{WH}+a_\mu^{\gamma h},
\label{twoloop MDM}
\end{eqnarray}
where $a_\mu^{WW},\;a_\mu^{WH},\;a_\mu^{\gamma h}$ are the contributions corresponding to Fig.\ref{Barzeediagrams}(a), (b), (c) respectively, and the concrete expressions can be found in Appendix \ref{au}. When we consider the PMNS mixing of neutrinos~\cite{PMNS1,PMNS2}, Fig.\ref{Barzeediagrams} (a) and (b) can also make contributions to the LFV processes. Hence, in the following numerical analysis, we also consider the contributions from the two loop Barr-Zee diagrams to the LFV processes.

\section{Numerical analyses\label{sec4}}
\indent\indent

In this section, we present the numerical results of the branching ratios for the LFV processes and muon MDM. The relevant SM input parameters are chosen as $m_W=80.385{\rm GeV},\;m_Z=90.1876{\rm GeV},\;\alpha_{em}(m_Z)=1/128.9,\;\alpha_s(m_Z)=0.118$. Since the tiny neutrino masses basically do not affect the numerical analysis, we take $Y_\nu=Y_x=0$ approximately. In addition, we consider the contributions from PMNS mixing to the LFV processes, then we take\cite{PDG}
\begin{eqnarray}
&&U_{PMNS}=
\left(\begin{array}{ccc} 0.8294,& 0.5391£¬ & 0.1462
\\ -0.4899,& 0.5764, & 0.6539
\\ 0.2682,& -0.6140, & 0.7422\end{array}\right).
\end{eqnarray}
The SM-like Higgs mass is\cite{PDG}
\begin{eqnarray}
&&m_h=125.09\pm0.24{\rm GeV}.
\end{eqnarray}
which constrains the corresponding parameter space strictly. In our previous work\cite{JLYang:2018}, we discussed the Higgs boson mass in the $B-L$ SSM in detail. Including the leading-log radiative corrections from stop
and top quark, we consider the constraint from the Higgs boson mass, hence our chosen parameter space in the following analysis satisfies the SM-like Higgs boson mass in experimental $3\sigma$ interval.

The updated experimental data~\cite{newZ} on searching $Z^\prime$ indicates $M_{Z^{'}}\geq4.05{\rm TeV}$ at 95\% Confidence Level (CL). Due to the contributions of heavy $Z'$ boson are highly suppressed, we choose $M_{Z'}=4.2{\rm TeV}$ in our following numerical analysis. And Refs.~\cite{20,21} give us an upper bound on the ratio between the $Z^{'}$ mass and its gauge coupling at 99\% CL as
\begin{eqnarray}
&&M_{Z^{'}}/g_{_B}\geq6{\rm TeV}\;.
\end{eqnarray}
then the scope of $g_{_B}$ is limited to $0<g_{_B}\leq0.7$. Additionally the LHC experimental data also constrain $\tan\beta^{'}<1.5$~\cite{8}. Considering the constraints from the experiments~\cite{PDG}, for those parameters in Higgsino and gaugino sectors, we appropriately fix $M_{1}=500{\rm GeV},\;M_{2}=600{\rm GeV},\;\mu=700{\rm GeV},\;\mu^{'}=800{\rm GeV},\;M_{BB^{'}}=500{\rm GeV},\;M_{BL}=600{\rm GeV}$, for simplify. For those parameters in the soft breaking terms, we take $B_\mu'=5\times10^5 {\rm GeV}^2$, $m_{\tilde{q}}=m_{\tilde{u}}=diag(2, 2, 1.6){\rm TeV}$, $T_u=diag(1, 1, 1){\rm TeV}$, to coincide with the constraints from the direct searches of squarks at the LHC\cite{ATLAS.PRD,CMS.JHEP} and the discussion about the observed Higgs signal in Ref.\cite{add1}. Considering the experiment observation on $\bar B\rightarrow X_s\gamma$ and $B_s^0\rightarrow \mu^+\mu^-$\cite{Yang:2018fvw}, we take $m_{H^\pm}=1.5{\rm TeV}$. All of the fixed parameters above do not affect the following numerical results obviously. Furthermore, in order to simplify our numerical analyses, we take soft breaking slepton mass matrices $m_{\tilde L,\tilde e}=diag(m_E, m_E, m_E){\rm TeV}$. But for the trilinear coupling matrix $T_e$ , we will introduce the slepton flavor mixing, which take into account the off-diagonal terms as
\begin{eqnarray}
&&T_e=
\left(\begin{array}{ccc} A_e,& \delta_{12} &\delta_{13}
\\ \delta_{12},&A_e, &\delta_{23}
\\ \delta_{13},&\delta_{23}, &A_e\end{array}\right){\rm TeV},
\end{eqnarray}

\subsection{Branching ratios for LFV processes}

\begin{figure}
\setlength{\unitlength}{1mm}
\centering
\includegraphics[width=3.1in]{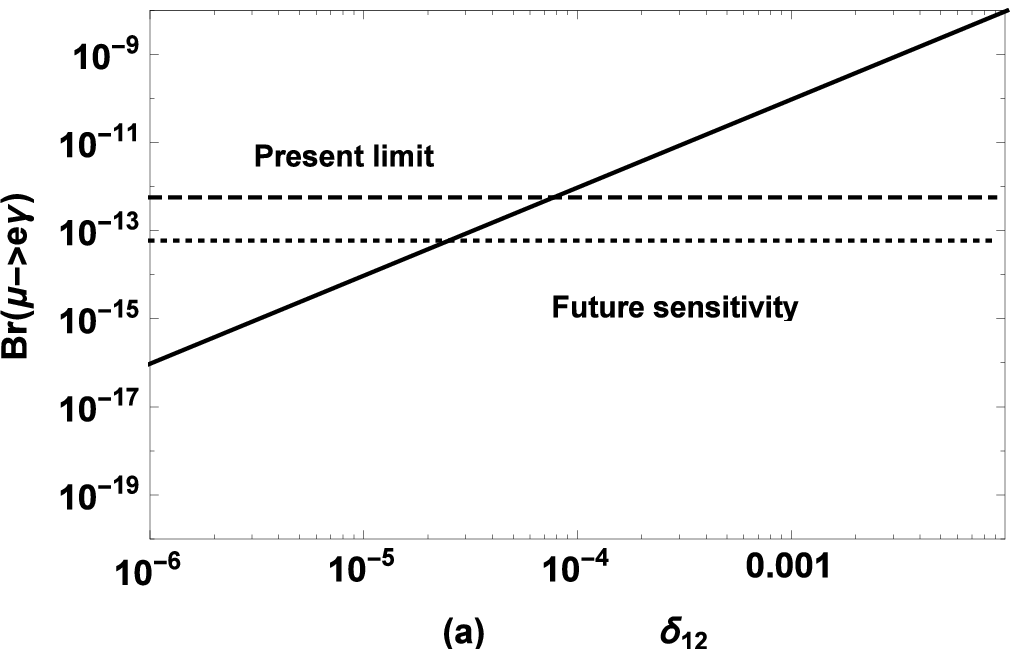}%
\vspace{0.5cm}
\includegraphics[width=3.1in]{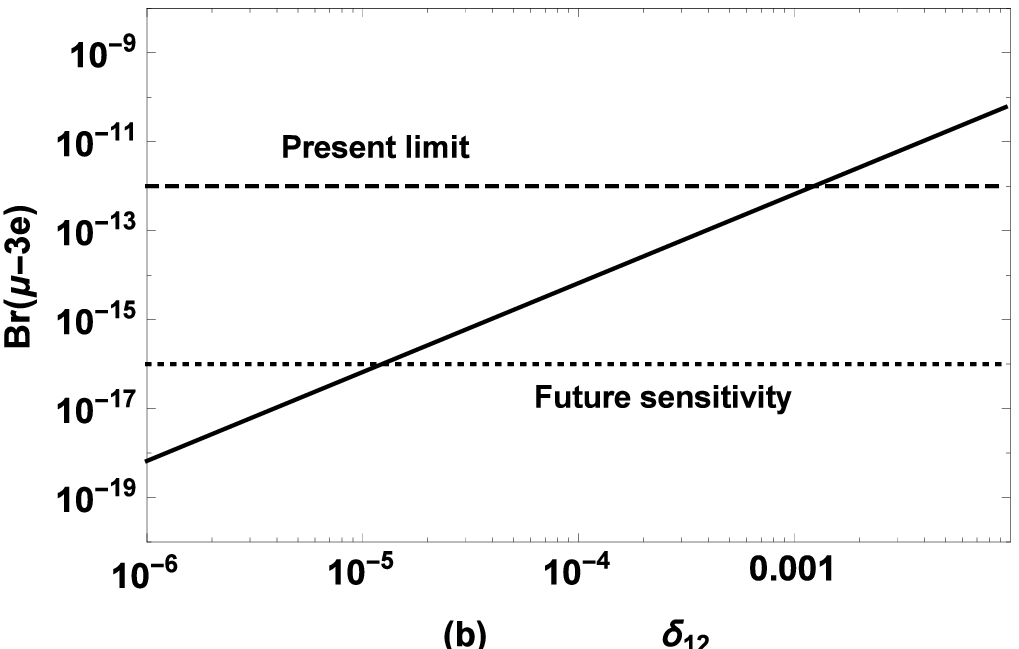}
\vspace{0cm}
\par
\hspace{-0.in}
\includegraphics[width=3.1in]{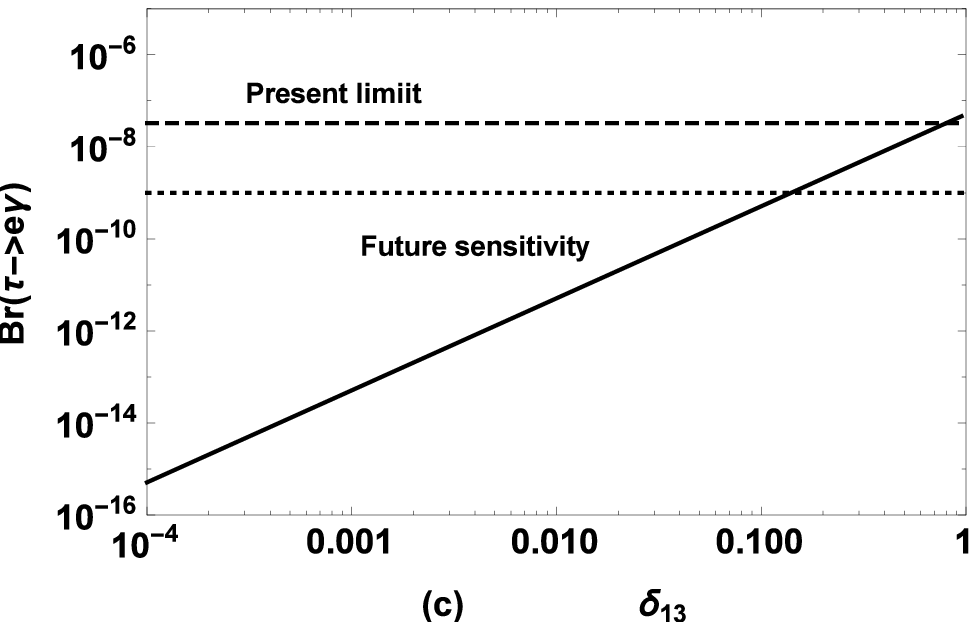}%
\vspace{0.5cm}
\includegraphics[width=3.1in]{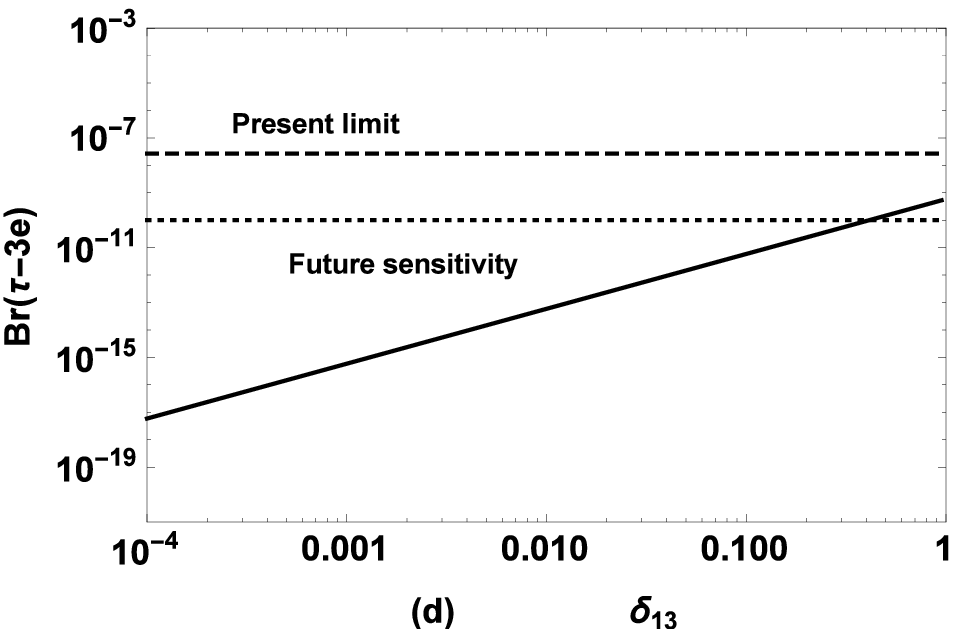}
\vspace{0cm}
\par
\hspace{-0.in}
\includegraphics[width=3.1in]{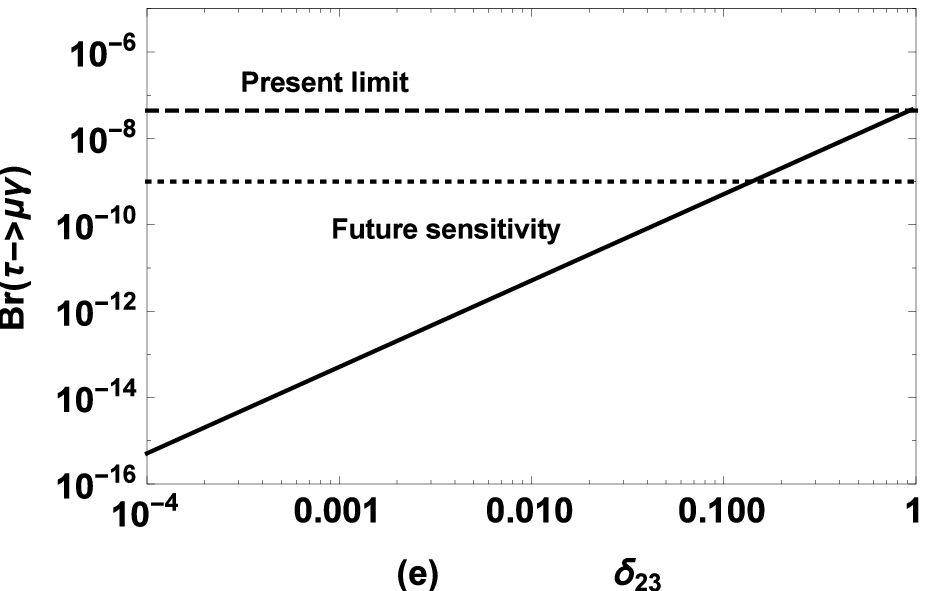}%
\vspace{0.5cm}
\includegraphics[width=3.1in]{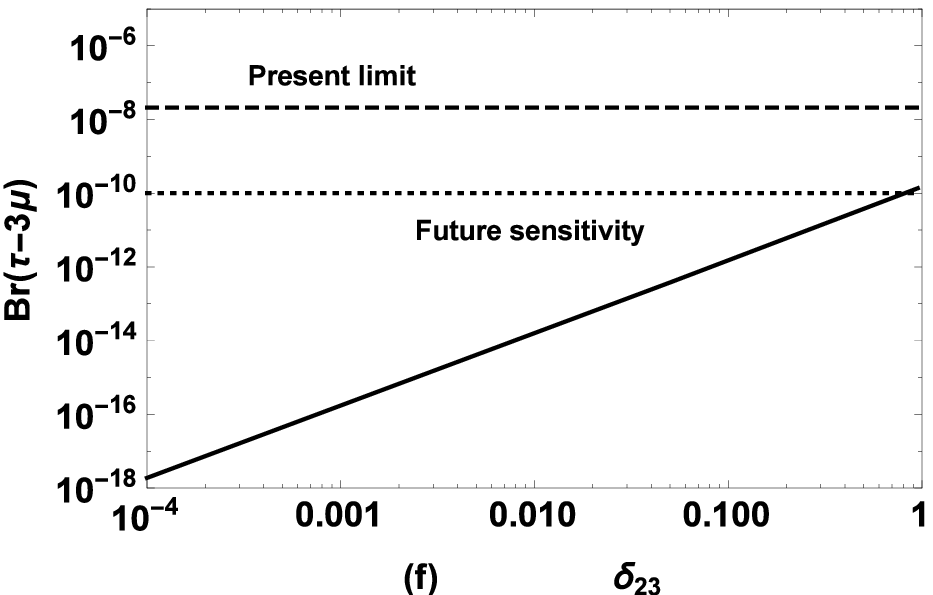}
\vspace{0cm}
\caption[]{LFV rates for $l_j-l_i$ transitions versus $\delta_{ij}$ are plotted, where the dashed and dotted lines denote the present limits and future sensitivities respectively.}
\label{Cljli}
\end{figure}

It is well known that the LFV branching ratios for $l_j-l_i$ transitions depend acutely on the mixing parameters $\delta_{ij}$. In order to see how $\delta_{ij}$ affect the theoretical evaluations of $l_j-l_i$ transitions, we assume that $\tan\beta=10,\;\tan\beta'=1.15,\;g_{_B}=0.4,\;g_{_{YB}}=-0.4,\;m_E=1,\;A_e=0.5$. Then we plot $Br(\mu\rightarrow e\gamma)$ and $Br(\mu\rightarrow 3e)$ versus $\delta_{12}$ for $\delta_{13}=\delta_{23}=0$ in Fig.\ref{Cljli} (a, b), in Fig.\ref{Cljli} (c, d) we picture $Br(\tau\rightarrow e\gamma)$ and $Br(\tau\rightarrow 3e)$ versus $\delta_{13}$ for $\delta_{12}=\delta_{23}=0$, $Br(\tau\rightarrow \mu\gamma)$ and $Br(\tau\rightarrow 3\mu)$ versus $\delta_{23}$ for $\delta_{12}=\delta_{13}=0$ are drawn in Fig.\ref{Cljli} (e, f), the dashed and dotted lines denote the present limits and future sensitivities respectively. It is obvious that the LFV rates increase with the increasing of slepton mixing parameters. Fig.\ref{Cljli} (a, b) shows that the present experimental limit bound of $Br(\mu\rightarrow e\gamma)$ constrains $\delta_{12}<10^{-4}$, which also coincides with the present experimental limit bound of $Br(\mu\rightarrow 3e)$. In addition, from Fig.\ref{Cljli} (c-f) we can see that $Br(\tau\rightarrow e\gamma)$ and $Br(\tau\rightarrow \mu\gamma)$ can reach the corresponding present experimental limit bounds, while $Br(\tau\rightarrow 3e)$ and $Br(\tau\rightarrow 3\mu)$ can't. However, the high future experimental sensitivities still keep a hope to detect $\tau\rightarrow 3e$ and $\tau\rightarrow 3\mu$. The two loop contributions are not obvious in Fig.\ref{Cljli}, because when $m_E=1{\rm TeV}$, the one loop results make the dominant contributions to the LFV processes, and the two loop results are negligible compared with the one loop results.

\begin{figure}[!h]
\setlength{\unitlength}{1mm}
\centering
\includegraphics[width=3.1in]{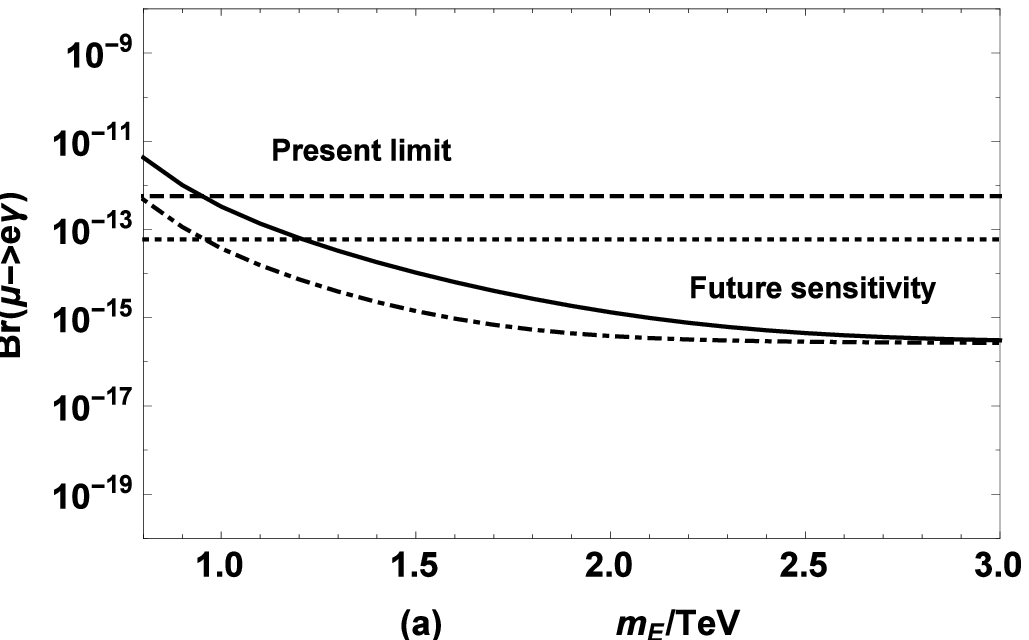}%
\vspace{0.5cm}
\includegraphics[width=3.1in]{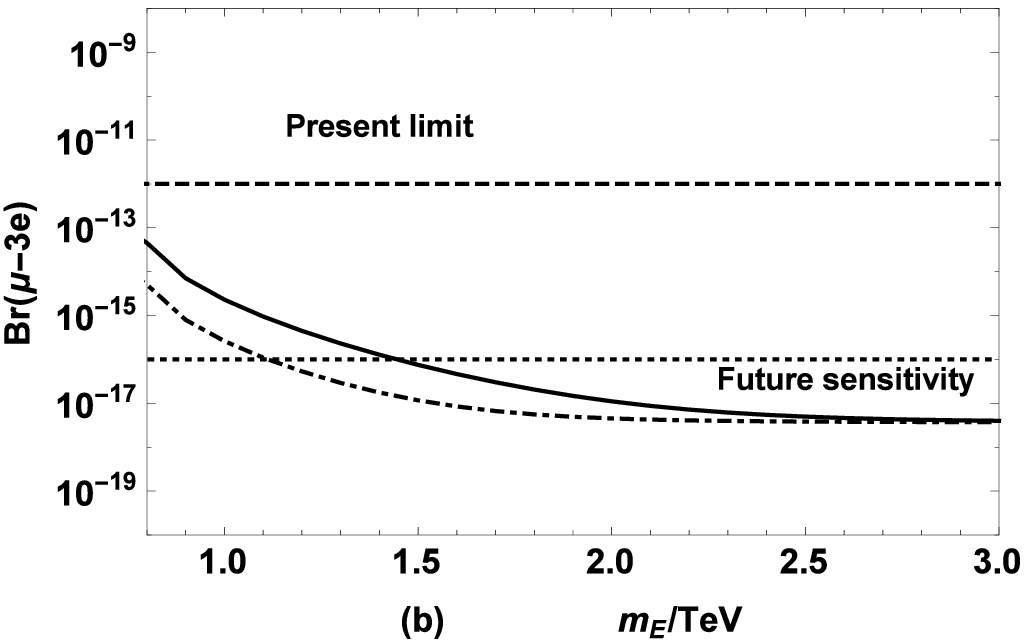}
\vspace{0cm}
\par
\hspace{-0.in}
\includegraphics[width=3.1in]{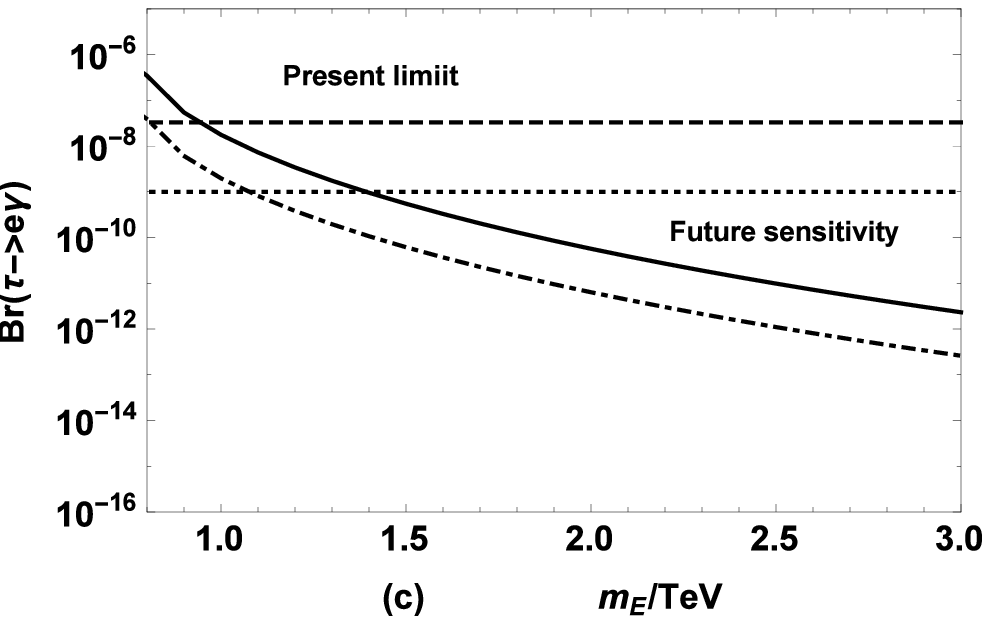}%
\vspace{0.5cm}
\includegraphics[width=3.1in]{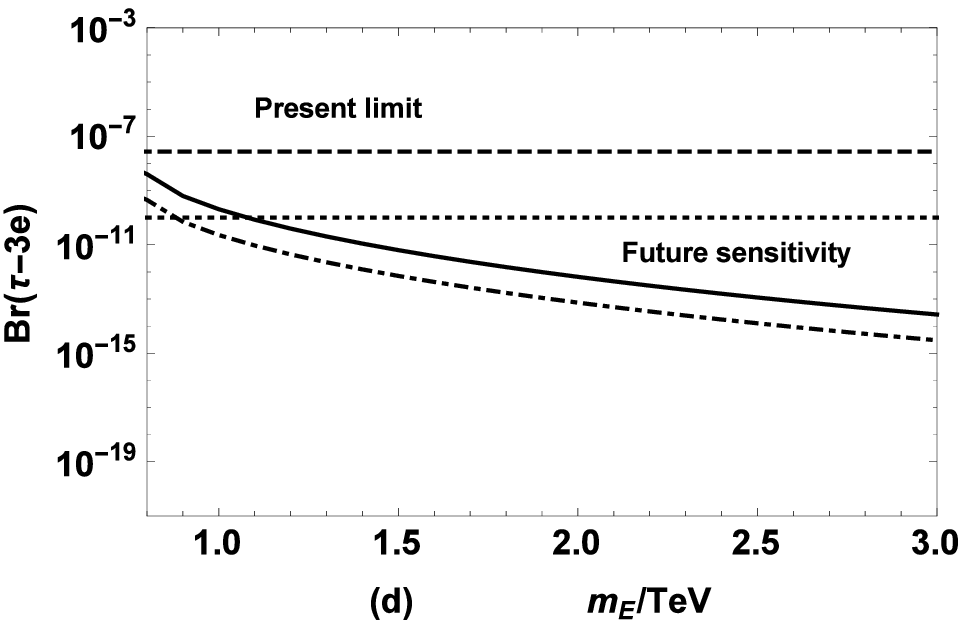}
\vspace{0cm}
\par
\hspace{-0.in}
\includegraphics[width=3.1in]{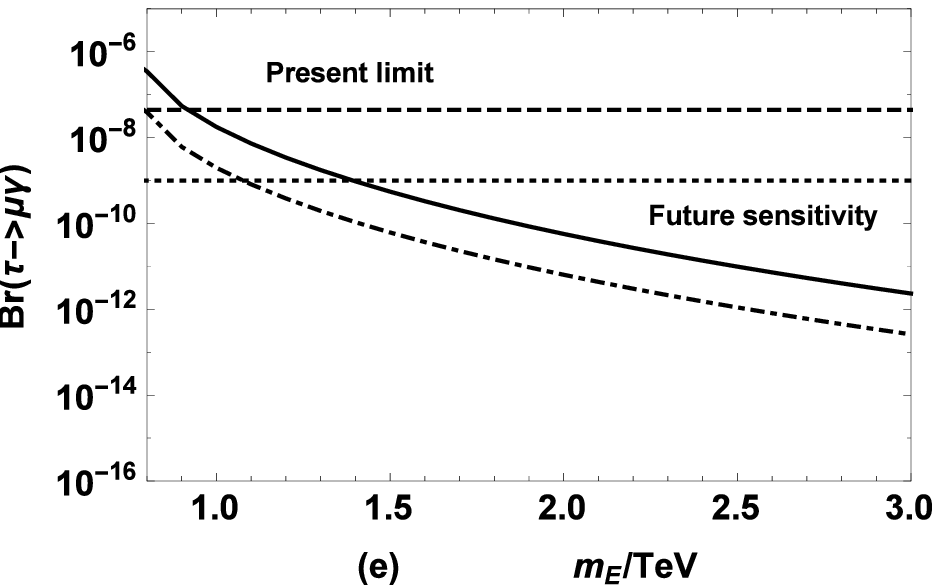}%
\vspace{0.5cm}
\includegraphics[width=3.1in]{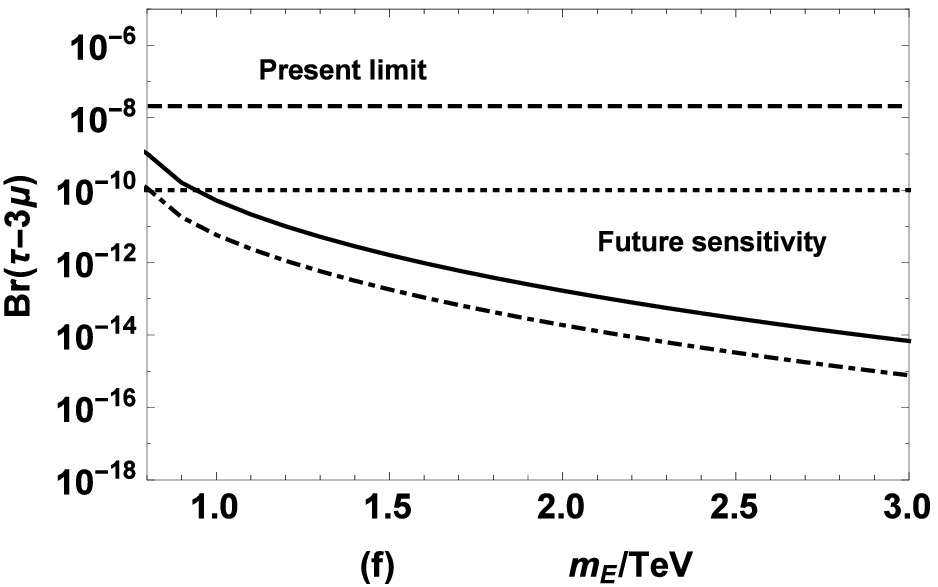}
\vspace{0cm}
\caption[]{LFV rates for $l_j-l_i$ transitions versus $m_E$ for $\tan\beta=10$ (solid lines), $\tan\beta=30$ (dotdashed lines) are plotted, where the dashed and dotted lines denote the present limits and future sensitivities respectively.}
\label{CljliME}
\end{figure}
Then we appropriately fix $\delta_{12}=5\times10^{-5},\;\delta_{13}=0.5,\;\delta_{23}=0.5$ to explore how the slepton masses and $\tan\beta$ affect the branching ratios for LFV transitions. We plot $Br(\mu\rightarrow e\gamma)$, $Br(\mu\rightarrow 3e)$, $Br(\tau\rightarrow e\gamma)$, $Br(\tau\rightarrow 3e)$, $Br(\tau\rightarrow \mu\gamma)$ and $Br(\tau\rightarrow 3\mu)$ versus $m_{E}$ for $\tan\beta=10$ (solid lines), $\tan\beta=30$ (dotdashed lines) in Fig.\ref{CljliME} (a-f), the dashed and dotted lines denote the present limits and future sensitivities respectively. It is obvious that the branching ratios for these processes decrease with the increasing of $m_E$ or $\tan\beta$, which indicates that heavy sleptons and large $\tan\beta$ play a suppressive role to the rates of LFV processes, and all of them can't reach the future sensitivities when $m_E\gtrsim1.5{\rm TeV}$. Fig.\ref{CljliME} (a, b) show that the numerical results depend on $m_E$ mildly when $m_E$ is large. Because the one loop contributions to $Br(\mu\rightarrow e\gamma)$, $Br(\mu\rightarrow 3e)$ are highly suppressed when $m_E$ is large and $\delta_{12}$ is small, then the two loop contributions can be comparable with the one loop results, and $m_E$ affects the two loop contributions negligibly. However, this feature does't appear in Fig.\ref{CljliME} (c-f), because when $\delta_{13},\;\delta_{23}$ are large, the two loop contributions to $Br(\tau\rightarrow e\gamma)$, $Br(\tau\rightarrow 3e)$, $Br(\tau\rightarrow \mu\gamma)$ and $Br(\tau\rightarrow 3\mu)$ are negligible compared with the one loop results. In addition, present experimental limit bounds of $Br(l_j\rightarrow l_i\gamma)$ constrain $m_E\gtrsim1{\rm TeV}$ for $\tan\beta=10$, and $Br(l_j\rightarrow l_i\gamma)$, $Br(l_j\rightarrow 3l_i)$ can reach the high future experimental sensitivities with small $m_E$.

\begin{table*}
\begin{tabular*}{\textwidth}{@{\extracolsep{\fill}}llll@{}}
\hline
parameters & min & max & step\\
\hline
$\tan\beta'$  & 1.02 & 1.5  & 0.01\\
$g_{_B}$      & 0.1  &  0.7 & 0.02\\
$g_{_{YB}}$   & -0.7 & -0.1 & 0.02\\
\hline
\end{tabular*}
\caption{Scanning parameters for Fig.\ref{Sljli}.}
\label{tab2}
\end{table*}

In order to see the effects of $\tan\beta',\;g_{_B},\;g_{_{YB}}$, which are new parameters in the $B-L$ SSM, we appropriately fix $\delta_{12}=5\times10^{-5},\;\delta_{13}=0.5,\;\delta_{23}=0.5,\;\tan\beta=10$ and $m_E=1$TeV. Then we scan the parameter space shown in Table \ref{tab2}.
\begin{figure}
\setlength{\unitlength}{1mm}
\centering
\includegraphics[width=3.1in]{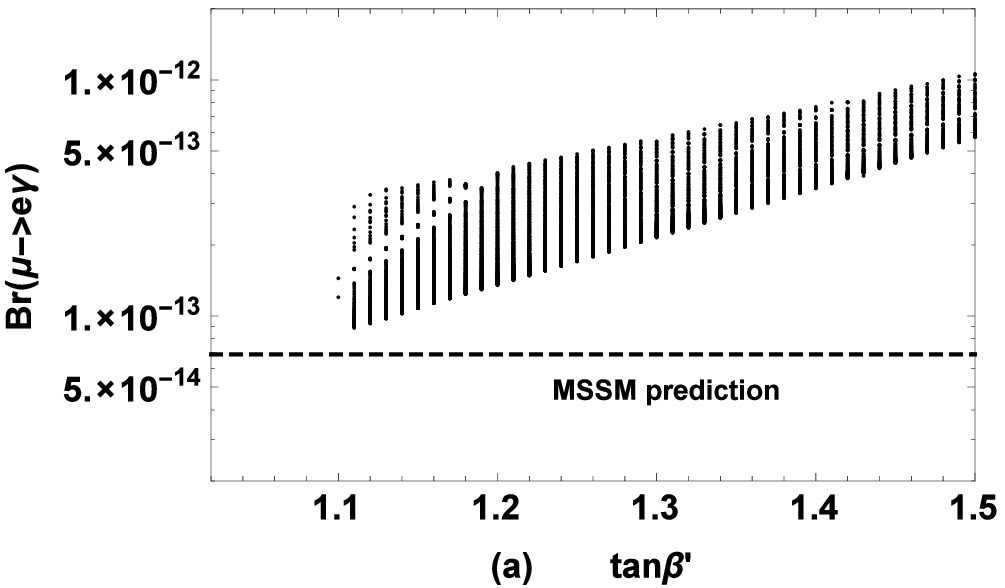}%
\vspace{0.5cm}
\includegraphics[width=3.1in]{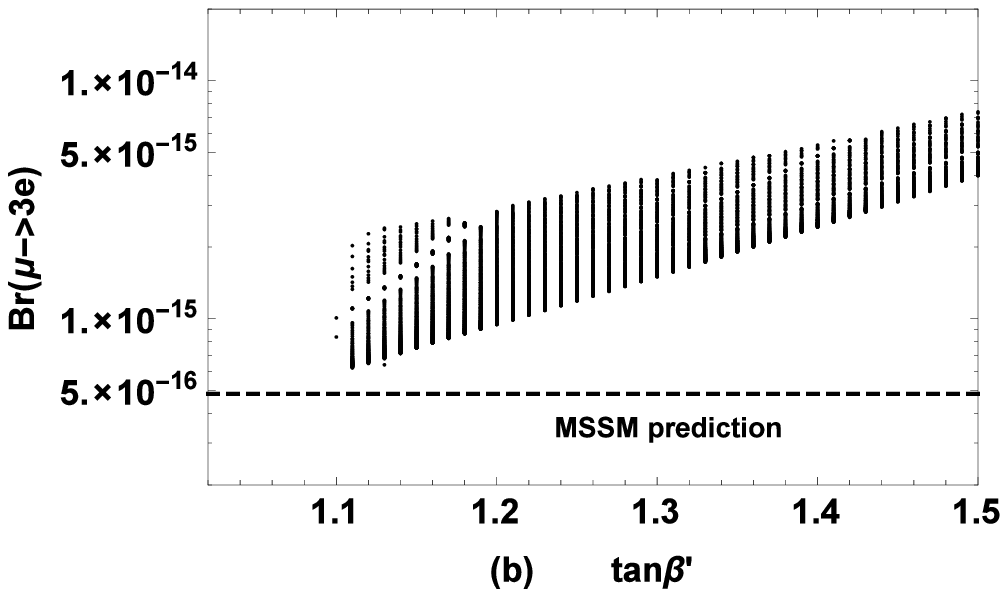}
\vspace{0cm}
\par
\hspace{-0.in}
\includegraphics[width=3.1in]{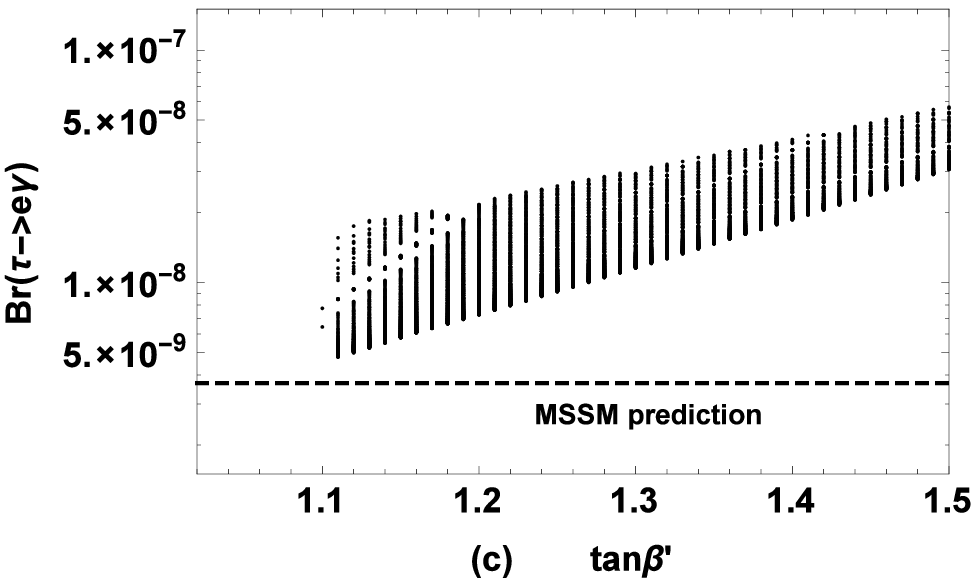}%
\vspace{0.5cm}
\includegraphics[width=3.1in]{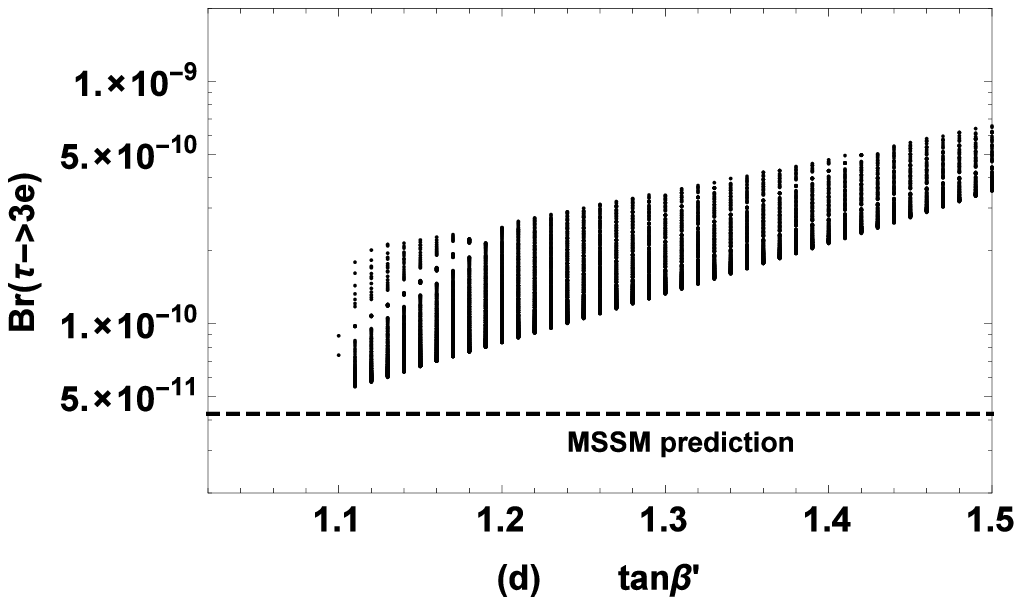}
\vspace{0cm}
\par
\hspace{-0.in}
\includegraphics[width=3.1in]{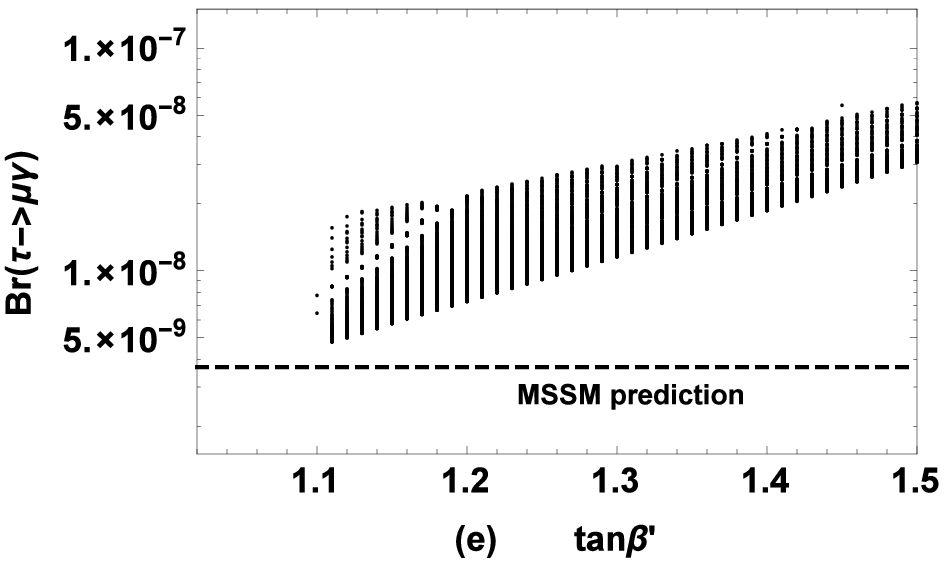}%
\vspace{0.5cm}
\includegraphics[width=3.1in]{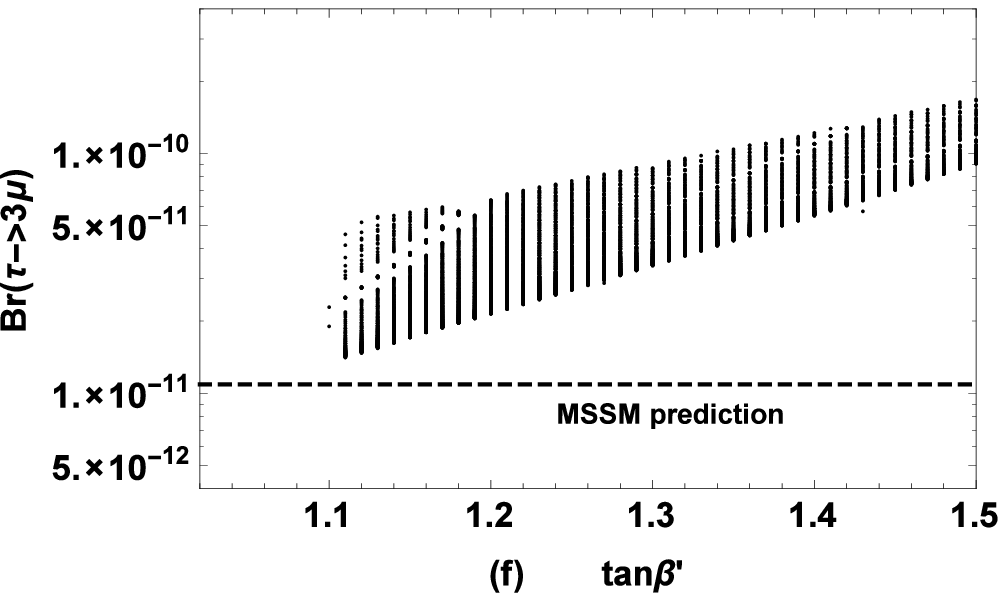}
\vspace{0cm}
\caption[]{LFV rates for $l_j-l_i$ transitions versus $\tan\beta'$ are plotted, where the dashed lines denote the MSSM prediction.}
\label{Sljli}
\end{figure}
In the scanning, we keep the slepton masses $m_{L_a}>500{\rm GeV} (a=1,\cdot\cdot\cdot,6)$, the Higgs boson mass in experimental $3\sigma$ interval, to avoid the range ruled out by the experiments\cite{PDG}. Then we plot $Br(\mu\rightarrow e\gamma)$, $Br(\mu\rightarrow 3e)$, $Br(\tau\rightarrow e\gamma)$, $Br(\tau\rightarrow 3e)$, $Br(\tau\rightarrow \mu\gamma)$ and $Br(\tau\rightarrow 3\mu)$ versus $\tan\beta'$ in Fig.\ref{Sljli} (a-f) respectively. In the same parameter space, the MSSM predicts that $Br(\mu\rightarrow e\gamma)\sim6.9\times10^{-14}$, $Br(\mu\rightarrow 3e)\sim4.8\times10^{-16}$, $Br(\tau\rightarrow e\gamma)\sim3.7\times10^{-9}$, $Br(\tau\rightarrow 3e)\sim4.2\times10^{-11}$, $Br(\tau\rightarrow \mu\gamma)\sim3.7\times10^{-9}$, $Br(\tau\rightarrow 3\mu)\sim1.1\times10^{-11}$. In order to see the differences between the $B-L$ SSM and MSSM predictions clearly, we also plot these MSSM predictions (dashed line) in Fig.\ref{Sljli} (a-f) respectively. The picture shows that the LFV rates increase with the increasing of $\tan\beta'$, and the numerical results depend on $\tan\beta',\;g_{_B},\;g_{_{YB}}$ comparably. When $\tan\beta'<1.1$, the range is excluded completely by concrete Higgs mass. In addition, it can be noted that all of the LFV rates can exceed the MSSM predictions easily. For example, in the $B-L$ SSM, $Br(\mu\rightarrow e\gamma)$ can reach $10^{-12}$ when $\tan\beta'=1.5$, which indicates that the new contributions in the $B-L$ SSM enhance the MSSM predictions about one order of magnitude. In Eq.(\ref{eq17}), we can see that the masses of sleptons decrease with the increasing of $\tan\beta'$ when $|g_{YB}|<g_B<2|g_{YB}|$. And the sleptons masses can decrease
from about $1000$GeV to $500$GeV with the increasing of $\tan\beta'$. In addition, since $\tan\beta',\;g_{_B},\;g_{_{YB}}$ affect the numerical results mainly through the new mass matrix of sleptons, and according to the decoupling theorem, we can conclude that large $\tan\beta'$ can enhance the theoretical predictions of these LFV processes when $|g_{YB}|<g_B<2|g_{YB}|$.

\subsection{Muon MDM}
Finally, we analyze the muon MDM in the $B-L$ SSM. Equation (\ref{aulimit}) shows that the NP contributions to the muon MDM should be constrained as $1.1\times10^{-10}<\Delta a_\mu^{NP}<48.5\times10^{-10}$, where we consider $3\sigma$ experimental error.

\begin{figure}
\setlength{\unitlength}{1mm}
\centering
\includegraphics[width=3.1in]{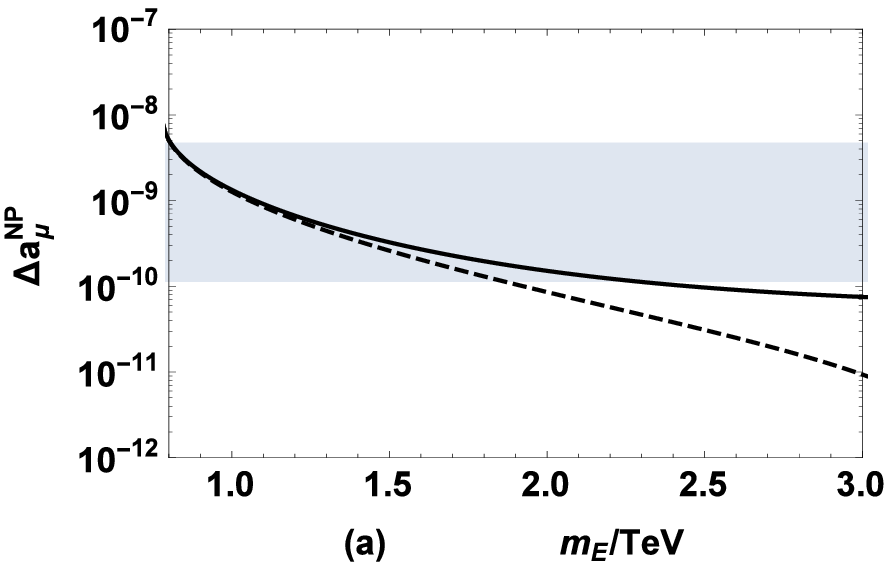}
\vspace{0.5cm}
\includegraphics[width=3.1in]{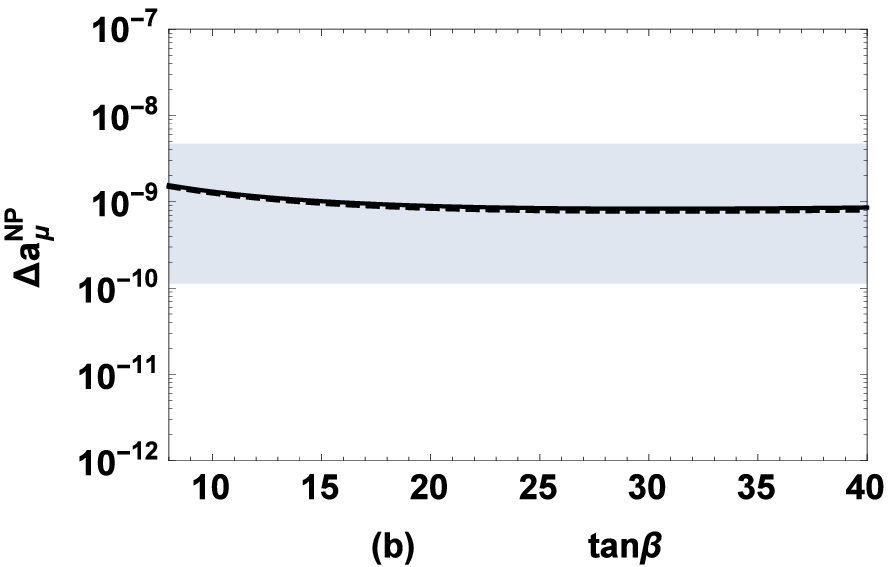}
\vspace{0cm}
\caption[]{$\Delta a_\mu^{NP}$ versus $m_E$ (a) and $\tan\beta$ (b) are plotted, where the solid and dashed line denote the two loop prediction, one loop prediction respectively, and the gray area denotes the experimental $3\sigma$ interval.}
\label{Camu}
\end{figure}

Taking $\tan\beta=10,\;\tan\beta'=1.15,\;g_{_B}=0.4,\;g_{_{YB}}=-0.4,\;A_e=0.5$, we plot the NP contributions to muon MDM in the $B-L$ SSM versus $m_E$ in Fig.\ref{Camu} (a). Then we take $m_E=1{\rm TeV}$ and plot $\Delta a_\mu^{NP}$ versus $\tan\beta$ in Fig.\ref{Camu} (b). Where the solid line denotes the two loop prediction, the dashed line represents the one loop prediction, and the gray area denotes the experimental $3\sigma$ interval.
Fig.\ref{Camu} (a) shows that $\Delta a_\mu^{NP}$ is decoupling with the increasing of $m_E$. The solid line and dashed line are separated more apparently with the increasing of $m_E$, which indicates that the one loop contributions are suppressed when $m_E$ is large, then the two loop results make the dominate contributions to $\Delta a_\mu^{NP}$. The main one-loop contributions to the muon MDM come from sleptons in Fig.~\ref{figllr}(a). And according to the decoupling theorem, the contributions from sleptons in Fig.~\ref{figllr} (a) and sneutrinos in Fig.~\ref{figllr} (b) are highly suppressed when $m_E$ is large enough. But sleptons and sneutrinos do not appear in the two-loop diagrams, hence the two-loop corrections to the muon MDM don't suffer such a suppressive factor, and the two-loop contributions can be dominant when $m_E$ is large enough. In addition, when the one loop contributions are highly suppressed, only two loop contributions can not reach the experimental $3\sigma$ bounds. However, the two loop diagrams also make important corrections to $\Delta a_\mu^{NP}$, hence we use the more precise two loop prediction in the following analysis. In Fig.\ref{Camu} (b) we can see that $\Delta a_\mu^{NP}$ decreases slowly with the increasing of $\tan\beta$, but $\tan\beta$ does not affect the numerical result obviously. And when $m_E=1{\rm TeV}$, one-loop corrections dominate the contributions to $\Delta a_\mu^{NP}$, hence the solid and dashed line almost appear as one in Fig.\ref{Camu} (b).

\begin{figure}
\setlength{\unitlength}{1mm}
\centering
\includegraphics[width=3.5in]{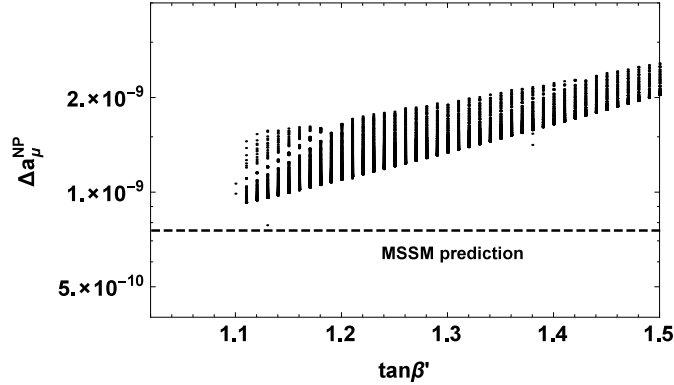}
\vspace{0cm}
\caption[]{$\Delta a_\mu^{NP}$ versus $\tan\beta'$ is plotted, where the dashed line denotes the MSSM prediction.}
\label{Samu}
\end{figure}

In order to see how $\tan\beta',\;g_{_B},\;g_{_{YB}}$ affect the theoretical prediction on $\Delta a_\mu^{NP}$, we take $m_E=1$ and scan the parameter space shown in Table \ref{tab2}. In the scanning, we also keep the slepton masses $m_{L_a}>500{\rm GeV} (a=1,\cdot\cdot\cdot,6)$, the Higgs boson mass in experimental $3\sigma$ interval. Then we plot $\Delta a_\mu^{NP}$ versus $\tan\beta'$ in Fig.\ref{Samu}.
In the same parameter space, the MSSM predicts that $\Delta a_\mu^{NP}\sim7.5\times10^{-10}$. In order to compare with the MSSM directly, we also plot the MSSM prediction (dashed line) in Fig.\ref{Samu}. The picture shows that, $\Delta a_\mu^{NP}$ increases with the increasing of $\tan\beta'$, which indicates that new parameters $\tan\beta',\;g_{_B},\;g_{_{YB}}$ also can affect the numerical result, and the effects of them are comparable. In addition, in the $B-L$ SSM, $\Delta a_\mu^{NP}$ can reach $2.6\times10^{-9}$ when $\tan\beta'$ is large. Since $\tan\beta',\;g_{_B},\;g_{_{YB}}$ affect the numerical result also mainly through the new mass matrix of sleptons, and the masses of sleptons decrease with the increasing of $\tan\beta'$ when $|g_{YB}|<g_B<2|g_{YB}|$, which implies that large $\tan\beta'$ can enhance the MSSM prediction on $\Delta a_\mu^{NP}$ when $|g_{YB}|<g_B<2|g_{YB}|$.

\section{Summary\label{sec5}}
\indent\indent
In this work, we focused on various LFV processes in the $B-L$ SSM with slepton flavor mixing, and analyze the two loop corrections to $\Delta a_\mu^{NP}$. Compared with the MSSM, new mass matrix of sleptons can affect the theoretical predictions on these processes. In addition, new $Z'$ gauge boson, new sneutrinos, new neutralinos and new Higgs bosons in the $B-L$ SSM can also make contributions. When the two loop corrections are included, new neutralinos can make contributions to $\Delta a_\mu^{NP}$ through the corresponding two loop diagrams. Considering the constraints from updated experimental data, in our chosen parameter space, the numerical results show that the present experimental limit bound of $Br(\mu\rightarrow e\gamma)$ constrains $\delta_{12}\lesssim10^{-4}$. In addition, all of these LFV rates decrease with the increasing of slepton masses or $\tan\beta$, and increase with the increasing of $\tan\beta'$ which is a new parameter in the $B-L$ SSM. The high future experimental sensitivities keep a hope to detect all of these LFV processes. In addition, the two loop diagrams make important corrections to $\Delta a_\mu^{NP}$. With respect to the MSSM, large $\tan\beta'$ can enhance the MSSM predictions on the branching ratios of LFV processes about one order of magnitude when $|g_{YB}|<g_B<2|g_{YB}|$, and also enhance the MSSM prediction on $\Delta a_\mu^{NP}$.

\begin{acknowledgments}
\indent\indent
The work has been supported by the National Natural Science Foundation of China (NNSFC) with Grants No. 11535002, No. 11647120, and No. 11705045, Natural Science Foundation of Hebei province with Grants No. A2016201010 and No. A2016201069, Hebei Key Lab of Optic-Eletronic Information and Materials, and the Midwest Universities Comprehensive Strength Promotion
project.
\end{acknowledgments}

\appendix

\section{The Wilson coefficients of the process $l_j^-\rightarrow l_i^-\gamma$. \label{wilsonllr}}
The coefficients corresponding to Fig.~\ref{figllr}(a), (b) can be written as
\begin{eqnarray}
&&A_1^{(a)L}=\frac{1}{6m_W^2}C_{\bar l_i F^n_k S^c_l}^L C_{\bar F^n_k S^{c*}_l l_j}^R I_4(x_{_{F^n_k}},x_{_{S^c_l}}),\nonumber\\
&&A_2^{(a)L}=\frac{m_{F^n_k}}{m_{l_j}m_W^2}C_{\bar l_i F^n_k S^c_l}^L C_{\bar F^n_k S^{c*}_l l_j}^L [I_3(x_{_{F^n_k}},x_{_{S^c_l}})-I_1(x_{_{F^n_k}},x_{_{S^c_l}})],\nonumber\\
&&A_{1,2}^{(a)R}=A_{1,2}^{(a)L}(L\leftrightarrow R),\nonumber\\
&&A_1^{(b)L}=\frac{1}{6m_W^2}C_{\bar l_i F^c_k S^n_l}^R C_{\bar F^c_k S^{n}_l l_j}^L [I_1(x_{_{F^c_k}},x_{_{S^n_l}})-2I_2(x_{_{F^c_k}},x_{_{S^n_l}})-I_4(x_{_{F^c_k}},x_{_{S^n_l}})],\nonumber\\
&&A_2^{(b)L}=\frac{m_{F^c_k}}{m_{l_j}m_W^2}C_{\bar l_i F^c_k S^n_l}^L C_{\bar F^c_k S^{n}_l l_j}^L [I_1(x_{_{F^c_k}},x_{_{S^n_l}})-I_2(x_{_{F^c_k}},x_{_{S^n_l}})-I_4(x_{_{F^c_k}},x_{_{S^n_l}})],\nonumber\\
&&A_{1,2}^{(b)R}=A_{1,2}^{(b)L}(L\leftrightarrow R).
\end{eqnarray}
where $x_i=m_i^2/m_W^2$, $C_{abc}^{L,R}$ denotes the constant parts of the interaction vertex about $abc$, which can be got through SARAH, and $a, b, c$ denote the interactional particles, and the concrete expressions for the functions $I_{1,2,3,4}$ and $G_{1,2,3,4}$ below can be found in Ref.\cite{Zhang:2013hva,Zhang:2013jva}.

\section{The Wilson coefficients of the process $l_j^-\rightarrow l_i^-l_i^-l_i^+$. \label{wilsonl3l}}
The coefficients corresponding to N-penguin contributions can be written as
\begin{eqnarray}
&&F^L=\frac{1}{2e^2}C_{\bar l_i F^n_k S^cl}^RC_{S^{c*}_lNS^c_\beta}^RC_{\bar F^n_k S^{c*}_\beta l_j}G_2(x_{F^n_k},
x_{S^c_\beta},x_{S^c_{l}})\nonumber\\
&&\qquad\quad+\frac{m_{F^c_k}m_{F^c_\alpha}}{e^2m_W^2}C_{\bar l_iS^n_\beta F^c_k}^RC_{\bar F^c_k N F^c_\alpha}^L
C_{\bar F^c_\alpha S^n_\beta l_j}^LG_1(x_{S^n_\beta},x_{F^c_k},x_{F^c_\alpha})\nonumber\\
&&\qquad\quad-\frac{1}{2e^2}C_{\bar l_iS^n_\beta F^c_k}^RC_{\bar F^c_k N F^c_\alpha}^RC_{\bar F^c_\alpha S^n_\beta l_j}^LG_2(x_{S^n_\beta},x_{F^c_k},x_{F^c_\alpha}),\nonumber\\
&&F^R=F^L({L\leftrightarrow R}),
\end{eqnarray}

The coefficients corresponding to box-type diagrams are
\begin{eqnarray}
&&B_1^L=\frac{m_{F^n_k}m_{F^n_\alpha}}{e^2m_W^2}G_3(x_{F^n_k},x_{F^n_\alpha},x_{S^c_\beta},x_{S^c_l})C_{\bar l_i S^c_l F^n_k}^LC_{\bar F^n_k S^{c*}_\beta l_j}^LC_{\bar l_i S^c_l F^n_\alpha}^RC_{\bar F^n_\alpha S^{c*}_\beta l_i}^R\nonumber\\
&&\qquad\quad+\frac{1}{2e^2m_W^2}G_4(x_{F^n_k},x_{F^n_\alpha},x_{S^c_\beta},x_{S^c_l})[C_{\bar l_i S^c_l F^n_k}^RC_{\bar F^n_k S^{c*}_\beta l_j}^LC_{\bar l_i S^c_\beta F^n_\alpha}^RC_{\bar F^n_\alpha S^{c*}_l l_i}^L\nonumber\\
&&\qquad\quad+C_{\bar l_i S^c_l F^n_k}^LC_{\bar F^n_k S^{c*}_\beta l_j}^RC_{\bar l_i S^c_l\beta F^n_\alpha}^RC_{\bar F^n_\alpha S^{c*}_l l_i}^L]+\frac{1}{2e^2m_W^2}G_4(x_{F^c_k},x_{F^c_\alpha},x_{S^n_\beta},x_{S^n_l})\nonumber\\
&&\qquad\quad\times C_{\bar l_i S^n_l F^c_k}^RC_{\bar F^c_k S^{n}_\beta l_j}^LC_{\bar l_i S^n_\beta F^c_\alpha}^RC_{\bar F^c_\alpha S^{n}_l l_i}^L,\nonumber\\
&&B_2^L=-\frac{m_{F^n_k}m_{F^n_\alpha}}{2e^2m_W^2}G_3(x_{F^n_k},x_{F^n_\alpha},x_{S^c_\beta},x_{S^c_l})C_{\bar l_i S^c_l F^n_k}^RC_{\bar F^n_k S^{c*}_\beta l_j}^RC_{\bar l_i S^c_\beta F^n_\alpha}^LC_{\bar F^n_\alpha S^{c*}_l l_i}^L\nonumber\\
&&\qquad\quad+\frac{1}{4e^2m_W^2}G_4(x_{F^n_k},x_{F^n_\alpha},x_{S^c_\beta},x_{S^c_l})[C_{\bar l_i S^c_l F^n_k}^RC_{\bar F^n_k S^{c*}_\beta l_j}^LC_{\bar l_i S^c_\beta F^n_\alpha}^LC_{\bar F^n_\alpha S^{c*}_l l_i}^R\nonumber\\
&&\qquad\quad+C_{\bar l_i S^c_l F^n_k}^RC_{\bar F^n_k S^{c*}_\beta l_j}^LC_{\bar l_i S^c_l F^n_\alpha}^RC_{\bar F^n_\alpha S^{c*}_\beta l_i}^L]+\frac{1}{4e^2m_W^2}G_4(x_{F^c_k},x_{F^c_\alpha},x_{S^n_\beta},x_{S^n_l})\nonumber\\
&&\qquad\quad\times C_{\bar l_i S^n_l F^c_k}^RC_{\bar F^c_k S^{n}_\beta l_j}^LC_{\bar l_i S^n_\beta F^c_\alpha}^LC_{\bar F^c_\alpha S^{n}_l l_i}^R-\frac{m_{F^c_k}m_{F^c_\alpha}}{2e^2m_W^2}G_3(x_{F^c_k},x_{F^c_\alpha},x_{S^n_\beta},x_{S^n_l})\nonumber\\
&&\qquad\quad\times C_{\bar l_i S^n_l F^c_k}^RC_{\bar F^c_k S^{n}_\beta l_j}^RC_{\bar l_i S^n_\beta F^c_\alpha}^LC_{\bar F^c_\alpha S^{n}_l l_i}^L,\nonumber\\
&&B_3^L=\frac{m_{F^n_k}m_{F^n_\alpha}}{e^2m_W^2}G_3(x_{F^n_k},x_{F^n_\alpha},x_{S^c_\beta},x_{S^c_l})[C_{\bar l_i S^c_l F^n_k}^LC_{\bar F^n_k S^{c*}_\beta l_j}^LC_{\bar l_i S^c_\beta F^n_\alpha}^LC_{\bar F^n_\alpha S^{c*}_l l_i}^L\nonumber\\
&&\qquad\quad-\frac{1}{2}C_{\bar l_i S^c_l F^n_k}^LC_{\bar F^n_k S^{c*}_\beta l_j}^LC_{\bar l_i S^c_l F^n_\alpha}^LC_{\bar F^n_\alpha S^{c*}_\beta l_i}^L]+\frac{m_{F^c_k}m_{F^c_\alpha}}{2e^2m_W^2}G_3(x_{F^c_k},x_{F^c_\alpha},x_{S^n_\beta},x_{S^n_l})\nonumber\\
&&\qquad\quad\times C_{\bar l_i S^n_l F^c_k}^RC_{\bar F^c_k S^{n}_\beta l_j}^RC_{\bar l_i S^n_\beta F^c_\alpha}^LC_{\bar F^c_\alpha S^{n}_l l_i}^L,\nonumber\\
&&B_4^L=\frac{m_{F^n_k}m_{F^n_\alpha}}{8e^2m_W^2}G_3(x_{F^n_k},x_{F^n_\alpha},x_{S^c_\beta},x_{S^c_l})C_{\bar l_i S^c_l F^n_k}^LC_{\bar F^n_k S^{c*}_\beta l_j}^LC_{\bar l_i S^c_l F^n_\alpha}^LC_{\bar F^n_\alpha S^{c*}_\beta l_i}^L,\nonumber\\
&&B_{1,2,3,4}^R=B_{1,2,3,4}^L({L\leftrightarrow R}).
\end{eqnarray}

\section{The SUSY contributions to the MDM of the muon. \label{au}}
The one loop contributions to MDM corresponding to Fig.~\ref{figllr}(a), (b) can be written as
\begin{eqnarray}
&&a_\mu^{(a)}=\Re\Big\{4x_\mu[-I_3(x_{F^n_k},x_{S^c_l})+I_4(x_{F^n_k},x_{S^c_l})][(C_{\bar\mu S^c_l F^n_k}^LC_{\bar F^n_k S^{c*}_l \mu}^R)+(C_{\bar\mu S^c_l F^n_k}^RC_{\bar F^n_k S^{c*}_l \mu}^L)^*]\nonumber\\
&&\qquad\quad+\sqrt{x_{\mu}x_{F^n_k}}[-I_3(x_{F^n_k},x_{S^c_l})+I_4(x_{F^n_k},x_{S^c_l})]C_{\bar\mu S^c_l F^n_k}^RC_{\bar F^n_k S^{c*}_l \mu}^R\Big\},\nonumber\\
&&a_\mu^{(b)}=\Re\Big\{x_\mu[-I_1(x_{F^c_k},x_{S^n_l})+2I_3(x_{F^c_k},x_{S^n_l})][(C_{\bar\mu S^n_l F^c_k}^RC_{\bar F^c_k S^n_l \mu}^L)+(C_{\bar\mu S^n_l F^c_k}^LC_{\bar F^c_k S^n_l \mu}^R)^*]\nonumber\\
&&\qquad\quad+\sqrt{x_{\mu}x_{F^c_k}}[2I_1(x_{F^c_k},x_{S^n_l})-2I_2(x_{F^c_k},x_{S^n_l})]C_{\bar\mu S^n_l F^c_k}^RC_{\bar F^c_k S^n_l \mu}^R\Big\}.
\end{eqnarray}

Under the assumption $m_F=m_{\chi^+_i}=m_{\chi^0_j}\gg m_W$, $m_F=m_{\chi^+_i}\gg m_{h}$, the Barr-Zee type diagrams contributing to the muon MDM corresponding to Fig.~\ref{Barzeediagrams}(a), (b), (c) can be simplify as
\begin{eqnarray}
&&a_\mu^{WW}=\frac{G_F m_\mu^2}{192\sqrt{2}\pi^4}\Big\{5(|C_{\bar\chi_j^0W^-\chi_i^+}^L|^2+|C_{\bar\chi_j^0W^-\chi_i^+}^R|^2)-
6(|C_{\bar\chi_j^0W^-\chi_i^+}^L|^2-|C_{\bar\chi_j^0W^-\chi_i^+}^R|^2)\nonumber\\
&&\qquad\quad+11\Re(C_{\bar\chi_j^0W^-\chi_i^+}^LC_{\bar\chi_j^0W^-\chi_i^+}^{R*})\Big\},\nonumber\\
&&a_\mu^{WH}=\frac{G_F m_\mu m_W^2 C_{\bar\mu H^- \nu_l}^L}{128\pi^4 m_F g_2}\Big\{\Big[\frac{179}{36}+\frac{10}{3}J(m_F^2,m_W^2,m_{H^\pm}^2)\Big]\Re(C_{\bar\chi_j^0W^-\chi_i^+}^L
C_{\bar\chi_j^0W^-\chi_i^+}^L\nonumber\\
&&\qquad\quad+C_{\bar\chi_j^0W^-\chi_i^+}^RC_{\bar\chi_j^0W^-\chi_i^+}^R)+\Big[-\frac{1}{9}-\frac{2}{3}
J(m_F^2,m_W^2,m_{H^\pm}^2)\Big]\Re(C_{\bar\chi_j^0W^-\chi_i^+}^LC_{\bar\chi_j^0W^-\chi_i^+}^R\nonumber\\
&&\qquad\quad+C_{\bar\chi_j^0W^-\chi_i^+}^RC_{\bar\chi_j^0W^-\chi_i^+}^L)+\Big[-\frac{16}{9}-\frac{8}{3}
J(m_F^2,m_W^2,m_{H^\pm}^2)\Big]\Re(C_{\bar\chi_j^0W^-\chi_i^+}^LC_{\bar\chi_j^0W^-\chi_i^+}^L\nonumber\\
&&\qquad\quad-C_{\bar\chi_j^0W^-\chi_i^+}^RC_{\bar\chi_j^0W^-\chi_i^+}^R)+\Big[-\frac{2}{9}-\frac{4}{3}
J(m_F^2,m_W^2,m_{H^\pm}^2)\Big]\Re(C_{\bar\chi_j^0W^-\chi_i^+}^LC_{\bar\chi_j^0W^-\chi_i^+}^R\nonumber\\
&&\qquad\quad-C_{\bar\chi_j^0W^-\chi_i^+}^RC_{\bar\chi_j^0W^-\chi_i^+}^L)\Big\},\nonumber\\
&&a_\mu^{\gamma h}=\frac{-G_F m_\mu m_W^2 C_{\bar\mu h^0 \mu}}{32\pi^4 m_F}\Re(C_{\bar\chi^+_i h^0 \chi^+_j}^L)\Big[1+\ln\frac{m_F^2}{m_h^2}\Big].
\end{eqnarray}
where
\begin{eqnarray}
&&J(x,y,z)=\ln x-\frac{y\ln y-z\ln z}{y-z}.
\end{eqnarray}

\end{document}